\documentclass[letterpaper,pre,twocolumn]{revtex4}

\usepackage{graphicx}
\usepackage{dcolumn}
\usepackage{amsmath}
\usepackage{mdwlist}

\newcommand{\etal}{\textit{et~al.}}
\newcommand{\e}{\mathrm{e}}
\newcommand{\dd}{\mathrm{d}}

\newcommand{\set}[1]{\lbrace#1\rbrace}

\newcommand{\Beta}{\mathop{\textrm{B}}}

\begin{document}

\title{Mixture models for data with unknown distributions}

\author{M. E. J. Newman}
\affiliation{Center for the Study of Complex Systems, University of Michigan, Ann Arbor, Michigan 48109, USA}

\begin{abstract}
  We describe and analyze a broad class of mixture models for real-valued multivariate data in which the probability density of observations within each component of the model is represented as an arbitrary combination of basis functions.  Fits to these models give us a way to cluster data with distributions of unknown form, including strongly non-Gaussian or multimodal distributions, and return both a division of the data and an estimate of the distributions, effectively performing clustering and density estimation within each cluster at the same time.  We describe two fitting methods, one using an expectation-maximization (EM) algorithm and the other a Bayesian non-parametric method using a collapsed Gibbs sampler.  The former is numerically efficient, but gives only point estimates of the probability densities.  The latter is more computationally demanding but returns a full Bayesian posterior and also an estimate of the number of components.  We demonstrate our methods with a selection of illustrative applications and give code implementing both algorithms.
\end{abstract}
\maketitle

\section{Introduction}
\label{sec:intro}
A mixture model describes data in which each observation or measurement belongs to one of several groups, or components, and the distribution of observations varies from one component to another~\cite{TSM85,MMR05,GMR22,MP00}.  Fits to mixture models allow us to divide data according to their distributions and are a powerful and widely used form of model-based data clustering.

In the most common version of this approach, the Gaussian mixture model, the distributions are Gaussians with unique means and variances in each component.  Many data, however, are not Gaussian, and mixture models have been developed for such data also, including models based on $t$ distributions, skew-normal and skew-$t$ distributions, beta and Dirichlet distributions, categorical distributions, or heterogeneous combinations of these, among many possibilities~\cite{McCutcheon87,MP00,AC03,Fruhwirth06,LLSL18,MHB16,MLR19}.  These models, however, still restrict the data to specific functional forms selected by the user.  In this paper we investigate an alternative class of models in which the observations within each component are drawn from an arbitrary distribution represented as a linear combination of a set of basis functions.  This formulation provides broad flexibility about the form of the distributions while remaining numerically tractable and practical for the analysis of real data.  It can encompass single-peaked distributions similar to normal or skew-normal distributions, but also a wide range of non-normal and multimodal distributions, all within the same family.

\begin{figure}
\begin{center}
\includegraphics[width=\columnwidth]{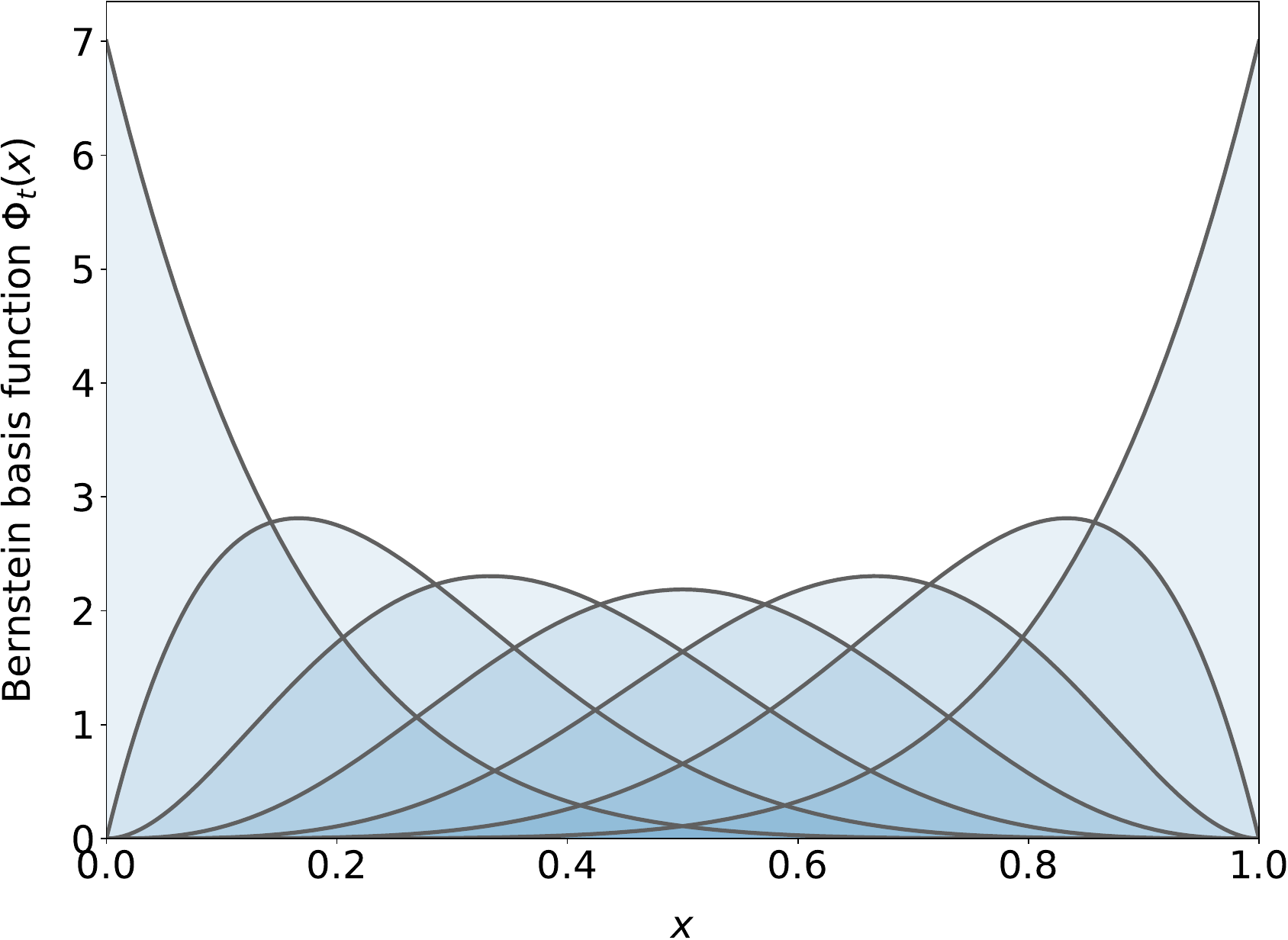}
\end{center}
\caption{The seven Bernstein basis functions of degree $d=6$, Eq.~\eqref{eq:bernstein}.}
\label{fig:bernstein}
\end{figure}

To give an example, one convenient choice of basis functions which we use in this paper is the \textit{Bernstein basis polynomials}~\cite{Lorenz86,Kakizawa04,Petrone99,WG19}.  The Bernstein basis of degree~$d$ is a set of $d+1$ polynomials $\Phi_t(x)$ defined on the domain $[0,1]$.  Each polynomial is of degree~$d$ and takes the form of a beta distribution with integer exponents:
\begin{equation}
\Phi_t(x) = {x^t (1-x)^{d-t}\over\Beta(t+1,d-t+1)}
  = {(d+1)!\over t!(d-t)!}\,x^t (1-x)^{d-t},
\label{eq:bernstein}
\end{equation}
for $t=0\ldots d$, where $\Beta(x,y)$ is the Euler beta function.  For example, the seven Bernstein basis functions for $d=6$ are shown in Fig.~\ref{fig:bernstein}.  Generically, they are non-negative single-peaked functions with equally spaced maxima.  The Bernstein basis is a complete basis for polynomials of degree~$d$---any such polynomial can be expressed as a linear combination---so our probability distributions are polynomials of specified degree, defined on the interval~$[0,1]$.

This is only one choice of basis.  Many others are possible, and for multivariate data our models allow different bases for different variables within the same model: both the number of basis functions and the functions themselves can change from one variable to another.  For instance, we might use a basis with a finite domain, such as the Bernstein basis, for one variable and a different basis on an unbounded domain for another variable.  We develop the theory in a general manner that allows any choice of basis for any variable.

An alternative way to interpret our models is as a \textit{mixture of mixtures}---a mixture model in which the data within each component are themselves generated from a mixture of elementary distributions~\cite{MP00,MH18}---but we depart from previous implementations of this idea, which have mainly focused on the case where the elementary distributions are Gaussians and one fits both the mixture of Gaussians and their parameters---means and variances---to the data.  In our formulation, by contrast, the basis functions are fixed and parameter-free---see the Bernstein basis of Eq.~\eqref{eq:bernstein} again for an example.  Only the mixture of functions is varied during the fit, not their shape.  Nonetheless, if the basis functions span a suitably wide selection of functional forms (such as polynomials of given degree), then the resulting model has the power to fit a correspondingly wide selection of data, while its simplicity leads to highly efficient numerical methods for fitting and sampling.

The relation to mixture-of-mixture models also highlights an important caveat of our approach.  Mixtures of mixtures are not normally applicable to \emph{univariate} data, because the sub-mixtures become completely non-identifiable: there is no way to distinguish based on the data alone between a mixture of mixtures and a standard one-level mixture model based on the same elementary distributions.  Only by applying rather elaborate priors can one impose a distinction between the two levels of the hierarchy~\cite{WB99,MFG17}.  The primary use case for the models of this paper, therefore, is in the analysis of multi\-variate data, for which the model parameters become identifiable, up to a permutation of the component labels.

This multivariate analysis can be thought of as a version of \textit{latent profile analysis}, a technique popular particularly in the social sciences, that clusters multivariate observations according to their probability distributions~\cite{BWP14,Oberski16,WK16,HF20,Spurk20,Bauer22}.  In the most common version of the method, one assumes just a single Gaussian distribution for each measured property, in which case the method becomes equivalent to a standard Gaussian mixture model and is of limited interest.  It becomes much more powerful, however, if the probability densities can take general and non-Gaussian forms.  This is what our models do.

To give a toy example, imagine measuring properties of a basket of assorted fruits, properties such as weight, diameter, hue, pH, and so forth.  Fitting these data with a mixture model might reveal that the fruits plausibly cluster into several groups, such that each group has its own characteristic distribution for each property, and further analysis might find that those groups in fact correspond to different fruit species---apples, oranges, etc.  In effect, the analysis would have inferred the existence of the species from their combined characteristics.  Note, however, that apples generally come in two hues, red and green, but not in any of the colors in between (orange, yellow), so the hue variable might have a bimodal distribution.  A Gaussian model would not be able to fit this distribution well: either the model would incorrectly infer that apples were two fruits, not one, or it would fail to fully make use of the information contained in the hue variable.  Our models, on the other hand, can address such situations and infer the correct structure from the data.  We give examples in Section~\ref{sec:results}.

\subsection{Previous work}
Previous work on mixture models with non-normal distribution functions has concentrated particularly on $t$ distributions, skew-normal and skew $t$-distributions~\cite{FP10,LM14,VM14}, and, for bounded data, beta~\cite{Ji05,ML09} and Dirichlet~\cite{BZV04} distributions (not to be confused with Dirichlet process mixture models).  Much of this work has focused on fitting algorithms, goodness-of-fit tests, and model selection.  Work on mixture-of-mixture models has primarily focused on the Gaussian case, which is effective for clustered data on unbounded domains, but is numerically demanding and limited in its ability to capture more general data distributions~\cite{DGR07,MFG17}.  Latent profile analysis, while promising in principle, has similarly focused almost exclusively on the Gaussian case, for which it reduces to an orthogonal Gaussian mixture with zero covariance~\cite{BWP14,Oberski16,WK16,HF20,Spurk20,Bauer22}.  Work on the non-normal case has been scarce~\cite{MHB16}.  Straddling the boundary between traditional mixture models and profile analysis is latent class analysis, the multivariate categorical cousin of the real-valued methods of this paper, about which much has been written~\cite{Goodman74,McCutcheon87,BR93,LLSL18}.  For latent class analysis, however, the model is unambiguous and no decision need be made about the distribution functions: the categorical (or Dirichlet-categorical) distribution is the only choice.

The Monte Carlo algorithm employed in this paper is a collapsed Gibbs sampler that directly samples component number as well as assignments of observations to components.  This algorithm owes a debt to previous methods for Dirichlet process models~\cite{Das14,Khoufache23} but, despite their elegance, and some previous discussion~\cite{BCG10,WWM16,NR16}, direct model selection approaches of this kind have rarely been used, perhaps because of poor equilibration when implemented using conventional sampling techniques~\cite{MH18}.  Proposed solutions for this problem, some of them quite complex, include the allocation sampler of Nobile and Fearnside~\cite{NF07} for Gaussian mixtures, the fast Gibbs sampler of Porteous~\etal~\cite{Porteous08} for latent Dirichlet allocation, and the split-merge sampler of Jain and Neal~\cite{JN04} for Dirichlet process models.  Reversible-jump Monte Carlo~\cite{Green95,RG97} takes a somewhat different approach, abandoning sampling from the marginal posterior in favor of a conventional parametric Gibbs sampler, modified to account for variation in the size of the parameter space when the number of components changes.  The approach we use, which employs techniques different from any of these, is an extension of one previously introduced for latent class analysis in~\cite{Newman25}.

\subsection{Our contributions}
In this paper we investigate the broad class of mixture models for multivariate real-valued data in which the data distributions within each component of the mixture are convex combinations of sets of fixed basis functions that we specify.  Fits to these models allow us to cluster observations and simultaneously estimate the distributions, which are interesting in their own right.  We describe two fitting methods, both applicable for any choice of basis functions.  The first is an expectation-maximization (EM) algorithm and the second is a non-parametric Bayesian Markov chain Monte Carlo.  The EM algorithm has the advantage of numerical efficiency, but requires that the user specify the number of components before performing the calculation.  The Monte Carlo approach returns a full Bayesian posterior on the component assignments and can also be used to infer the number of components when the latter is unknown.  It is more computationally demanding than the EM algorithm but fast enough for convenient interactive application to data sets of typical size.  We present a series of illustrative applications of our methods to data including both synthetic (computer-generated) benchmarks and real-world examples.  Complete code implementing our methods in Python and C is available at \verb|https://umich.edu/~mejn/lpa| for download.

\section{The model}
\label{sec:model}
Consider a data set consisting of $N$ multivariate observations, for each of which we measure $M$ real-valued items or variables.  For example, the items might be answers to questions on a survey or results of a set of medical tests, recorded for $N$ individuals.  The measured value for individual~$i$ of item~$j$ we denote by $x_{ij}$ with $i=1\ldots N$ and $j=1\ldots M$ and we assume the data to be complete---there are no missing measurements.

We assume the values to be generated by a mixture model in which each observation belongs to one of $k$ components or profiles, with different distributions of observations in each component.  The defining feature of our model is that the probability density of each observed value~$x$ is drawn not from any specific distribution, such as a Gaussian, but from an arbitrary distribution represented in parametric form as a combination of a set of basis functions, with a separate such combination for each component/item pair.  Specifically, let $\Phi_{jt}(x)$ with $t=0\ldots T_j-1$ be a set of $T_j$ non-negative basis functions defined over any convenient domain and normalized over that domain such that
\begin{equation}
\int \Phi_{jt}(x) \>\dd x = 1 \qquad\mbox{for all $j$ and $t$.}
\label{eq:norm}
\end{equation}
Crucially, the basis functions are fixed and parameter-free, and are not fitted to the data.  Only the \emph{amount} of each function in the combination is fitted.  This differs from typical mixture-of-mixture models based on Gaussians, for example, for which the parameters of the individual Gaussians (mean, variance) are fitted to the data~\cite{MFG17}.  (Having $t$ run from $0\ldots T_j-1$, rather than $1\ldots T_j$ which might seem more natural, is not strictly necessary, but all the basis sets we use in our own calculations start from zero so it is a convenient choice.)

Now consider the convex combination
\begin{equation}
f(x) = \sum_{t=0}^{T_j-1} \theta_t \Phi_{jt}(x),
\label{eq:fx}
\end{equation}
for some set of parameters $\theta_t\ge0$ with $\sum_t \theta_t = 1$.  This combination is non-negative, since each of the individual basis functions is non-negative, and is unit normalized over the domain of interest since
\begin{equation}
\int f(x) \>\dd x = \sum_{t=0}^{T_j-1} \theta_t \int \Phi_{jt}(x) \>\dd x
  = \sum_{t=0}^{T_j-1} \theta_t = 1,
\end{equation}
where we have used Eq.~\eqref{eq:norm}.  This makes the family of functions~\eqref{eq:fx} ideal for representing probability densities, and this is the form we use for the profiles in our model.  For an observation~$i$ in component~$r$ we write the probability of a measured value~$x_{ij}$ for item~$j$ as
\begin{equation}
P(x_{ij}|r,\theta_{rj}) = \sum_{t=0}^{T_j-1} \theta_{rjt} \Phi_{jt}(x_{ij}),
\label{eq:pxrj}
\end{equation}
for some parameters~$\theta_{rjt}$, with $\theta_{rj}$ denoting the set for $t=0\ldots T_j-1$.  In fitting our model we will infer not only the assignments of individuals to components but also the values of these parameters, which in turn gives us an estimate of the distributions from which the data in each component are drawn.

We will not assume that the set of basis functions is the same for all items~$j$.  In some cases they may be, but other data sets may benefit from using a different representation for different items.  For instance, some items may have bounded values while others are unbounded.  Some may have values concentrated in particular regions while others are more uniform.  Some may have more complicated distributions that benefit from representations with a larger number of basis functions, and so forth.  We develop the theory for the general case in which each item~$j$ has its own number~$T_j$ of basis functions and its own functions~$\Phi_{jt}(x)$.

An alternative way to interpret the model is as a mixture of mixtures, in which the distribution of Eq.~\eqref{eq:pxrj} is generated from another mixture model, with components indexed by~$t$, which we will refer to as \textit{slots}.  An observation~$x$ is generated by first drawing a slot $t=0\ldots T_j-1$ from a categorical distribution with probabilities~$\theta_{rjt}$ and then drawing the observed value~$x$ from the basis function for that slot.  Then
\begin{equation}
P(x_{ij}|r,\theta_{rj}) = \!\sum_{t=0}^{T_j-1} P(t|r,\theta_{rj}) P(x_{ij}|t)
  = \!\sum_{t=0}^{T_j-1} \theta_{rjt} \Phi_{jt}(x_{ij}),
\label{eq:pxrjtheta}
\end{equation}
which recovers Eq.~\eqref{eq:pxrj}.  In our calculations, however, we will delay performing the sum over~$t$, modeling instead the full joint distribution over components~$r$ and slots~$t$.  Defining $g_i$ to be the component to which $i$ belongs and $h_{ij}$ to be the slot from which the value~$x_{ij}$ is drawn, we have $P(x_{ij},h_{ij}|g_i,\theta_{rj}) = \theta_{g_ijh_{ij}} \Phi_{jh_{ij}}(x_{ij})$ and, assuming independence, the likelihood of all values and slots is
\begin{equation}
P(x,h|g,\theta) = \prod_{ij} \theta_{g_ijh_{ij}} \Phi_{jh_{ij}}(x_{ij})
  = \prod_{ij} \theta_{g_ijh_{ij}} \phi_{ijh_{ij}},
\label{eq:likelihood}
\end{equation}
where we have introduced the shorthand
\begin{equation}
\phi_{ijt} = \Phi_{jt}(x_{ij}).
\label{eq:phi}
\end{equation}
This formulation will be crucial to the efficient numerical implementation of our models.

We will further assume \textit{a~priori} that, for given~$k$, individual~$i$ is assigned to a component~$g_i$ with categorical probabilities $\pi = (\pi_1,\ldots,\pi_k)$ which we will estimate as part of our procedure, assuming a uniform (uninformative) Dirichlet prior.  (It is possible to fit the model with general, non-uniform Dirichlet priors as well---it is a straightforward extension of the process described here---but for all the calculations presented in this paper we make the uniform choice.)  We also assume a uniform Dirichlet prior on the parameters~$\theta_{rj}$.  Then the full posterior probability of all model variables is
\begin{align}
P(g,h,\pi,\theta|x,k) &= {P(x,h|g,\theta) P(\theta) P(g|\pi) P(\pi|k)\over P(x)}
    \nonumber\\
   &= \prod_i \Bigl[ \pi_{g_i} \prod_j \theta_{g_ijh_{ij}} \phi_{ijh_{ij}} \Bigr],
\label{eq:pfull}
\end{align}
where we have neglected some multiplicative constants in the second line that play no role in further developments.

Note that the quantities~$\phi_{ijt}$ of Eq.~\eqref{eq:phi} are constant, independent of any of the parameters of the model, and hence need be evaluated only once at the start of the calculation.  This evaluation is the only time the basis functions~$\Phi_{jt}(x)$ enter into our calculations and the choice of basis plays no further role once the $\phi_{ijt}$ have been computed.  In practice, this means that we are agnostic about which basis functions we use and we are at liberty to make any convenient choice, or to make different choices for different items if we wish.  All subsequent developments in this paper require only the values~$\phi_{ijt}$ and are equally applicable for any basis or bases.  The same is also true of computer code implementing our methods.  The code takes as input only the values~$\phi_{ijt}$ and does not need to know either the data $x_{ij}$ or the basis functions from which the values are derived.  This means that the exact same code can be used, without modification, to fit the model with any choice of basis.  In Appendix~\ref{app:basis} we give a selection of possible basis functions for use with various types of data, including data with bounded, semi-infinite, and infinite domains, piecewise constant densities, and densities defined on the circle.

\section{Fitting the model}
\label{sec:fitting}
In this section we describe two approaches for fitting the model of Section~\ref{sec:model} to data.  The first is an EM algorithm, which provides point estimates of the probability distributions for each component/item combination and a posterior distribution over the component assignments, conditioned on the probability distributions.  The second approach is a Bayesian non-parametric Monte Carlo algorithm that returns an unconditional posterior on the assignments as well as an estimate of the number of components, but at the expense of greater computational effort.

\subsection{EM algorithm}
\label{sec:em}
The first of our two proposed fitting methods is an expectation-maximization (EM) algorithm~\cite{DLR77,MK08}.  For the purposes of this algorithm we assume that the number of components~$k$ is known---this method cannot be used to select the value of~$k$, although in Section~\ref{sec:bayesian} we describe an alternative method that can and other methods for selecting $k$ could also be applied, based for instance on the Bayesian information criterion~\cite{RD06}.

Our goal with the EM algorithm is to make a maximum a posteriori (MAP) estimate of the parameters~$\pi,\theta$ and then use that estimate to infer the values of the component assignments~$g_i$.  The derivation of the algorithm is a standard one.  The only subtlety is that we compute the joint distribution over both components and slots and not just the components.  The slots are not normally of interest, but including them allows us to write formulas in closed form, which otherwise is not possible.

Summing Eq.~\eqref{eq:pfull} over $g_i = 1\ldots k$ for all~$i$ and $h_{ij}=0\ldots T_j-1$ for all~$i,j$, we write the marginal posterior probability of the parameters as
\begin{align}
P(\pi,\theta|x) &= \sum_{\set{g_i}} \sum_{\set{h_{ij}}} P(g,h,\pi,\theta|x) \nonumber\\
  &= \sum_{\set{g_i}} \sum_{\set{h_{ij}}}
     \prod_i \Bigl[ \pi_{g_i} \prod_j \theta_{g_ijh_{ij}} \phi_{ijh_{ij}} \Bigr],
\label{eq:ppitheta}
\end{align}
where we have suppressed $k$ in the notation, since it will remain constant throughout the developments.  We want to maximize this expression with respect to both $\pi$ and~$\theta$, or equivalently (and more easily) maximize its logarithm
\begin{equation}
\log P(\pi,\theta|x)
  = \log \sum_{\set{g_i}} \sum_{\set{h_{ij}}} P(g,h,\pi,\theta|x).
\label{eq:em1}
\end{equation}
Direct maximization is difficult, so instead we apply Jensen's inequality, which says that for any positive quantities~$x_i$
\begin{equation}
\log \sum_i x_i \ge \sum_i q_i \log {x_i\over q_i},
\label{eq:jensen}
\end{equation}
where~$q_i$ is any set of non-negative weights such that $\sum_i q_i = 1$.  Applied to Eq.~\eqref{eq:em1}, this implies that
\begin{equation}
\log P(\pi,\theta|x)
  \ge \sum_{\set{g_i}} \sum_{\set{h_{ij}}}  q\bigl(\set{g_i},\set{h_{ij}}\bigr)
      \log {P(g,h,\pi,\theta|x)\over q\bigl(\set{g_i},\set{h_{ij}}\bigr)},
\label{eq:ineq}
\end{equation}
where we can regard $q\bigl(\set{g_i},\set{h_{ij}}\bigr)$ as a probability distribution with $\sum_{\set{g_i}} \sum_{\set{h_{ij}}} q\bigl(\set{g_i},\set{h_{ij}}\bigr) = 1$, and the exact equality is recovered when
\begin{align}
q\bigl(\set{g_i},\set{h_{ij}}\bigr) &= {P(g,h,\pi,\theta|x)\over
    \sum_{\set{g_i}} \sum_{\set{h_{ij}}} P(g,h,\pi,\theta|x)} \nonumber\\
  &= {P(g,h,\pi,\theta|x)\over P(\pi,\theta|x)}.
\label{eq:estep}
\end{align}
Together, Eqs.~\eqref{eq:ineq} and~\eqref{eq:estep} tell us that if we maximize the right-hand side of~\eqref{eq:ineq} with respect to~$q\bigl(\set{g_i},\set{h_{ij}}\bigr)$, it will become equal to the left-hand side, and if we maximize \emph{that} result with respect to the parameters~$\pi,\theta$, we will get our desired MAP estimate.  On its face this seems to have made the problem more complicated, turning what was originally a simple maximization of Eq.~\eqref{eq:em1} into a joint maximization with respect to both the parameters and the new quantity~$q\bigl(\set{g_i},\set{h_{ij}}\bigr)$.  In fact, however, it is a useful elaboration because it gives us a simple numerical way to perform the maximization: from any starting point we maximize repeatedly and alternately with respect~$q\bigl(\set{g_i},\set{h_{ij}}\bigr)$, which we do using Eq.~\eqref{eq:estep}, and with respect to the parameters.  Iterating this procedure to convergence then gives us our MAP estimate of the parameters.

To maximize with respect to the parameters we proceed as follows.  Using Eq.~\eqref{eq:pfull}, the terms on the right-hand side of~\eqref{eq:ineq} that depend on $\pi$ and $\theta$ are
\begin{align}
&\sum_{\set{g_i}} \sum_{\set{h_{ij}}} q\bigl(\set{g_i},\set{h_{ij}}\bigr)
  \Bigl[ \sum_i \log \pi_{g_i} +
  \sum_{ij} \log \theta_{g_ijh_{ij}} \Bigr] \nonumber\\
  &= \sum_{\set{g_i}} \sum_{\set{h_{ij}}} q\bigl(\set{g_i},\set{h_{ij}}\bigr)
     \Bigl[ \sum_i \sum_r \delta_{g_ir} \log \pi_r \nonumber\\
  &\hspace{14em} + \sum_{ij} \sum_{rt} \delta_{g_ir} \delta_{h_{ij}t}
       \theta_{rjt} \Bigr] \nonumber\\
  &= \sum_{ir} q^i_r \log \pi_r + \sum_{ijrt} q^{ij}_{rt} \log \theta_{rjt},
\label{eq:freeenergy}
\end{align}
where $\delta_{ij}$ is the Kronecker delta and
\begin{align}
q^i_r &= \sum_{\set{g_i}} \sum_{\set{h_{ij}}} q\bigl(\set{g_i},\set{h_{ij}}\bigr)
         \delta_{g_ir}, \\
q^{ij}_{rt} &= \sum_{\set{g_i}} \sum_{\set{h_{ij}}}
   q\bigl(\set{g_i},\set{h_{ij}}\bigr) \delta_{g_ir} \delta_{h_{ij}t}.
\end{align}
Physically, $q^i_r$ represents the marginal probability within the distribution $q\bigl(\set{g_i},\set{h_{ij}}\bigr)$ that individual~$i$ is assigned to component~$r$ and $q^{ij}_{rt}$ represents the marginal probability that individual~$i$ is assigned to group~$r$ and to slot~$t$ for item~$j$.  Using Eqs.~\eqref{eq:pfull} and~\eqref{eq:estep}, many factors now cancel and we find that
\begin{align}
\label{eq:qir}
q^i_r &= {\pi_r \prod_j \sum_{t=0}^{T_j-1} \theta_{rjt} \phi_{ijt}\over
  \sum_s \pi_s \prod_j \sum_{t=0}^{T_j-1} \theta_{sjt} \phi_{ijt}}, \\
\label{eq:qijrt}
q^{ij}_{rt} &= {\pi_r \theta_{rjt} \phi_{ijt}
               \prod_{k(\ne j)} \sum_{u=0}^{T_k-1} \theta_{rku} \phi_{iku}\over
               \sum_s \pi_s \prod_k \sum_{u=0}^{T_k-1} \theta_{sku} \phi_{iku}} \nonumber\\
  &= {\theta_{rjt} \phi_{ijt}\over\sum_{t=0}^{T_j-1} \theta_{rju} \phi_{iju}}\,q^i_r.
\end{align}
Armed with these quantities, we can now maximize~\eqref{eq:freeenergy} with respect to the parameters by simple differentiation.  Maximizing with respect to~$\pi_r$ and enforcing the constraint $\sum_r \pi_r = 1$ with a Lagrange multiplier, we find that
\begin{equation}
\pi_r = {1\over N} \sum_i q^i_r.
\label{eq:mstep1}
\end{equation}
Similarly maximizing with respect to $\theta_{rjt}$ and enforcing $\sum_t \theta_{rjt} = 1$ with another multiplier gives
\begin{equation}
\theta_{rjt} = {\sum_i q^{ij}_{rt}\over\sum_i q^i_r},
\label{eq:mstep2}
\end{equation}
where we have made use of the fact that $\sum_t q^{ij}_{rt} = q^i_r$ for all~$j$.

From any initial condition, such as a random assignment of the parameter values, our EM algorithm consists of applying Eqs.~\eqref{eq:qir} and~\eqref{eq:qijrt} to calculate~$q^i_r$ and $q^{ij}_{rt}$, then applying~\eqref{eq:mstep1} and~\eqref{eq:mstep2} to calculate $\pi_r$ and~$\theta_{rjt}$, and repeating until the values of all quantities converge.  Given these values, we then wish to estimate the component assignments themselves, but in fact this is already done.  From Eq.~\eqref{eq:estep} we have
\begin{equation}
q\bigl(\set{g_i},\set{h_{ij}}\bigr) = {P(g,h,\pi,\theta|x)\over P(\pi,\theta|x)}
  = P(g,h|\pi,\theta,x).
\end{equation}
In other words, the converged value of~$q\bigl(\set{g_i},\set{h_{ij}}\bigr)$ is precisely the joint posterior on the component assignments~$g$ and slot assignments~$h$, given the estimated parameter values.  Thus, for instance, the final value of~$q^i_r$, Eq.~\eqref{eq:qir}, gives us the posterior probability that individual~$i$ belongs to component~$r$.  If we want a single ``best'' component assignment, we could then assign each observation to the component it is most likely to belong to.  And the estimated probability densities from which the observations in each component are drawn can be reconstructed from Eq.~\eqref{eq:pxrj} and the inferred values of the~$\theta_{rjt}$.

As with all EM algorithms, this one is guaranteed to converge to a local maximum of the posterior probability, but whether it converges to the global maximum depends on the shape of the probability surface and the starting values of the parameters.  In practice it is wise to perform multiple runs to test for consistent convergence.

\subsection{Bayesian estimation}
\label{sec:bayesian}
The EM algorithm of the previous section works well and is numerically efficient, but it does have some disadvantages.  Like all EM algorithms it provides only point estimates of the parameters~$\pi,\theta$, which is acceptable when the data are sufficiently dense that the error on those estimates is small, but can be problematic for sparser data.  Point estimates also pose a risk of overfitting in sparse data sets, since the number of parameters can be very large for some calculations.  For instance, an application with $k=10$ components, $M=20$ items, and $T_j=5$ basis functions for each item would have 1000 parameters~$\theta_{rjt}$, which could easily approach the number of observations.

These issues can be remedied by taking a Bayesian approach to fitting the model, in which we sample from the complete posterior over component assignments using a Markov chain Monte Carlo method.  This provides a robust fitting mechanism that does not rely on point estimates.  Its main disadvantages are the inherent statistical error that accompanies all sampling-based methods and that it can be more computationally demanding than the EM algorithm, although, as we will see, calculations remain very tractable for data sets of typical size.

Our approach roughly follows the one outlined for general mixture models in~\cite{Newman25}.  Consider the following Bayesian version of our model, which can be thought of as an extension of the ``mixture of finite mixtures'' model of Miller and Harrison~\cite{MH18}.  First, we select the number of components~$k$ from some prior distribution~$P(k)$ that we choose.  The simplest choice is a uniform (uninformative) prior.  The minimum value of $k$ is~1 (all observations in a single component) and the maximum is~$N$ (every observation in its own component), so the uniform prior takes the form
\begin{equation}
P(k) = {1\over N}.
\label{eq:uniform}
\end{equation}
Many other choices are possible, however, and our method is agnostic about which one we use.  In our own calculations we use the uniform choice above, but we develop the theory for the general case.

Given the value of~$k$, the natural next step would be to select a component~$g_i$ for each observation~$i$ from the categorical distribution with parameters~$\pi_r$.  Then we integrate out the parameters, applying a flat Dirichlet prior as before, which gives
\begin{equation}
P(g|k) = {(k-1)!\over(N+k-1)!} \prod_{r=1}^k n_r!
\label{eq:standard1}
\end{equation}
where $n_r$ is the number of observations in component~$r$.  Another, perhaps quicker way to reach the same result, is to observe that the above procedure simply chooses the sizes~$n_r$ of the components uniformly at random from all possible choices such that $\sum_r n_r = N$.  The number of such choices is ${N+k-1\choose k-1}$ and the number of possible assignments of observations to components of the chosen sizes is the multinomial $N!/\prod_r n_r!$.  Hence the probability of any one assignment is
\begin{equation}
P(g|k) = {N+k-1\choose k-1}^{\!-1}\,{\prod_{r=1}^k n_r!\over N!},
\end{equation}
which reproduces~\eqref{eq:standard1}.

This is the approach taken by Miller and Harrison, but it is not ideal in our case because it can generate empty components.  Arguably this might be acceptable in situations where $k$ is fixed, but in the present context, where $k$ will be a free random variable, it makes little sense.  One proposed solution is to use a sparse model with a prior that favors empty components and then count only non-empty ones~\cite{VWRM15,FM19}, but we take a simpler approach: as suggested in~\cite{Newman25} we simply forbid empty components.  We select the component sizes uniformly from choices that sum to~$N$ as before, but considering only choices with non-empty components.  There are ${N-1\choose k-1}$ such choices,~so
\begin{equation}
P(g|k) = {N-1\choose k-1}^{\!-1} {\prod_r n_r!\over N!}.
\label{eq:pzk}
\end{equation}
This is the choice we make in this paper.

Having assigned the observations to components, we now select the slot from which each item~$j$ will be drawn from the categorical distribution with parameters~$\theta_{rjt}$.  If $h_{ij}$ is the slot for $i$'s measurement of item~$j$ as before, then the probability of the complete set~$h=\set{h_{ij}}$ of all slot assignments is
\begin{equation}
P(h|g,k,\theta) = \prod_{ij} \theta_{g_ijh_{ij}}
  = \prod_{rjt} \theta_{rjt}^{m_{rjt}},
\end{equation}
where $m_{rjt}$ is the number of observations in component~$r$ whose value for item~$j$ was drawn from slot~$t$.  Once again we impose a flat Dirichlet prior on~$\theta_{rj}$ and integrate, which gives
\begin{align}
P(h|g,k) &= \int P(h|g,k,\theta) P(\theta) \>\dd\theta \nonumber\\
  &= \prod_{rj} (T_j-1)! \int \prod_{t=0}^{T_j-1} \theta_{rjt}^{m_{rjt}} \>\dd\theta_{rj} \nonumber\\
  &= \prod_{rj} {(T_j-1)!\over(n_r+T_j-1)!} \prod_{t=0}^{T_j-1} m_{rjt}!
\end{align}

Finally, given the~$h_{ij}$, the likelihood of the observed data is
\begin{equation}
P(x|h) = \prod_{ij} \Phi_{jh_{ij}}(x_{ij}) = \prod_{ij} \phi_{ijh_{ij}}.
\end{equation}
Putting everything together, we have,
\begin{align}
P&(k,g,h|x) = {P(x|h) P(h|g,k) P(g|k) P(k)\over P(x)} \nonumber\\
  &= P(k) (k-1)! (N-k)! \nonumber\\
  &\quad\times \prod_r \biggl[ n_r!
     \prod_j \biggl( {(T_j-1)!\over (n_r+T_j-1)!}
     \prod_{t=0}^{T_j-1} m_{rjt}! \biggr) \biggr] \prod_{ij} \phi_{ijh_{ij}},
\label{eq:posterior}
\end{align}
where we have neglected some constants in the second line that play no role in our calculations.  Note that the factor of~$\prod_j (T_j-1)!$, although it might seem to be constant, cannot be neglected because the number of times it appears is proportional to~$k$, and $k$ will be allowed to vary in our calculations.

\subsection{Monte Carlo algorithm}
\label{sec:mc}
Our goal is to sample states $(k,g,h)$ from the marginal posterior~\eqref{eq:posterior}, which we do by sampling values for one observation at a time, holding the others constant, in the manner of a Gibbs sampler.  The broad approach is described for general mixture models in~\cite{Newman25}.  For the present model it consists of repeatedly performing the following steps.

\begin{enumerate*}
\setlength{\itemsep}{4pt}
\item Choose a component~$r$ uniformly at random.
\item Choose an observation~$i$ uniformly at random from component~$r$.
\item Remove $i$ from component~$r$.
\item If $i$ was the sole observation in component~$r$, so that its removal leaves the component empty, delete the component, relabel component~$k$ to be the new component~$r$, and decrease $k$ by~1.
\item Construct a set of $k+1$ candidate states.
\begin{enumerate*}
\item Of these, $k$~states are ones in which $i$ is placed in an existing component~$s = 1\ldots k$.  (We explicitly include the case $s=r$ where $i$ is placed back in the component it was just removed from, in which case its component assignment does not change, but its slot assignments may still change in step~7 below.)  Each of these $k$ states has an associated weight
\begin{equation}
w_s = {N-k\over k} P(k) \prod_j {\sum_{t=0}^{T_j-1}
       (m_{sjt}+1)\phi_{ijt}\over n_s+T_j},
\label{eq:w1}
\end{equation}
where $m_{sjt}$ and $n_s$ are the values after $i$ is removed from $r$ but before it is placed in its new home.
\item The last candidate state is one in which $i$ becomes the sole member of a new component~$k+1$.  The weight for this state is
\begin{equation}
w_{k+1} = k P(k+1) \prod_j {1\over T_j} \sum_{t=0}^{T_j-1} \phi_{ijt},
\label{eq:w2}
\end{equation}
where $k$ is the number of components before the new component is created.
\end{enumerate*}
\item Select a component~$s$ in the range $1\ldots k+1$ at random with categorical probabilities
\begin{equation}
p(s) = {w_s\over\sum_r w_r}.
\end{equation}
Assign $i$ to the component selected.  If this is a new component, increase $k$ by~1.
\item Assign $i$ a new slot~$t$ for each item~$j$ independently at random from a categorical distribution with probabilities
\begin{equation}
p_j(t) = {(m_{sjt}+1) \phi_{ijt}\over \sum_{u=0}^{T_j-1} (m_{sju}+1) \phi_{iju}}
\label{eq:slots1}
\end{equation}
in the general case, or
\begin{equation}
p_j(t) = {\phi_{ijt}\over\sum_{u=0}^{T_j-1} \phi_{iju}}
\label{eq:slots2}
\end{equation}
for a newly created component.
\end{enumerate*}
Repeating these steps will, from any starting state, cause the system to converge to the probability distribution of Eq.~\eqref{eq:posterior}, from which one can then draw samples $(k,g,h)$ of the number of components and component and slot assignments for each observation.  Note how in the first step of the algorithm we draw a random component, not a random observation, a crucial distinction that implements the rejection-free sampling from the prior proposed in~\cite{Newman25}, which overcomes the tendency of standard collapsed Gibbs samplers to undersample small components~\cite{MH18} without requiring the use of more complex remedies such as the split-merge method of Jain and Neal~\cite{JN04}.  Another feature of the algorithm is the way it selects components and slots separately in steps~6 and~7.  A naive sampler would select both in a single operation, but this is computationally intensive because of the combinatorially large number of component/slot pairings.  By separating the two operations one achieves the same outcome with much less effort.  A proof of the correctness of the algorithm is given in Appendix~\ref{app:mc}.

Implementation of the algorithm is straightforward, but there are a few technical tricks worth noting.  It is inefficient to implement the algorithm directly in terms of the component labels~$g_i$.  Instead, it is better to maintain a list, for instance in an array, of the members of each component, which is updated when members are added or removed and which allows one to quickly select a random member as required in step~2.  This approach also improves the efficiency of Monte Carlo moves that require exchanging the labels of entire components.  Instead of exchanging all the individual labels, we simply exchange pointers to the membership arrays.  The slot assignments~$h_{ij}$ do not change when components are relabeled, since they do not depend on the component labels.  Also, as mentioned earlier, the quantities~$\phi_{ijt}$ are constant throughout the calculation and need be evaluated only once at the start of the algorithm, then stored for later use.  And for some choices of basis functions, some of the formulas for the algorithm can be simplified.  For the Bernstein basis of Eq.~\eqref{eq:bernstein}, for example, we have $\sum_t \phi_{ijt} = d+1$ for all $i,j$, which simplifies Eqs.~\eqref{eq:w2} and~\eqref{eq:slots2}.

With these optimizations the algorithm runs quickly.  Our implementation performs up to about 10 million Monte Carlo steps per second on conventional hardware (\textit{circa} 2025).  In keeping with previous work, we measure the length of runs in ``sweeps,'' where a sweep means $N$ individual Monte Carlo steps, so that the component assignment of each observation gets updated once on average per sweep.  Monte Carlo algorithms require an initial equilibration phase or ``burn-in,'' in which one performs some number of sweeps to allow the probability distribution over states to converge to the correct posterior.  In our studies we use a burn-in of anywhere from 2500 to 10\,000 sweeps, followed by 25\,000 to 100\,000 sweeps for sampling, which is adequate to achieve good sample statistics and consistent results in most cases.  For typical data sets of the kind we consider in this paper, such runs take anywhere from a few seconds to a few minutes on standard hardware.

\subsection{Output of the algorithm}
The output of the algorithm is a sequence of samples of component/slot assignments for each observation, drawn from the marginal posterior distribution of the model, Eq.~\eqref{eq:posterior}, along with the number of components~$k$.  However, while such a set of samples accurately reflects the  posterior, it is often not the kind of output that a practitioner would find most useful.  In practice, one is more often looking for a single definitive assignment of observations to components.  The situation is further complicated by the phenomenon of \textit{label switching}, common to all such Monte Carlo methods~\cite{Stephens00a}.  The labels used to identify the components have no absolute significance: any permutation of the labels gives the same division of observations and hence all such permutations have equal probability under the posterior distribution.  Any Monte Carlo algorithm that correctly samples from the posterior will sample all permutations with equal frequency, which means that every observation belongs to each of the $k$ components with the same probability~$1/k$ on average and one cannot meaningfully assign any observation to a specific component---the labels are not identifiable.

A number of solutions to this problem have been advanced.  Some authors advocate for symmetry-breaking rules that favor certain permutations of the labels over others~\cite{RG97,Stephens00a}.  As we argue in~\cite{Newman25}, however, these approaches have technical problems that cause them to give biased results in some cases.  We prefer instead to use the complete sampled set of component assignments to create a \textit{consensus clustering} of the observations, which aims to summarize the broad features of the entire set in a single assignment.  This may not always be possible---the set may be so diverse as to defy such a summary~\cite{KN22}---but when the distribution over assignments is relatively concentrated, there are a number of methods that will generate a useful consensus clustering in reasonable time~\cite{MPMG03,Bryant03,GF08,VR11,LF12,ZDA23}.

Our Monte Carlo algorithm does not directly generate inferred values for the parameters~$\theta_{rjt}$ that describe the probability densities, Eq.~\eqref{eq:pxrj}, from which the observations are drawn, but given the sampled values of the component assignments~$g_i$ and slot assignments~$h_{ij}$, one can write
\begin{align}
P(\theta|x,g,h) &= {P(x,h|g,\theta)P(\theta)\over P(x) P(h)} \nonumber\\
  &= {1\over P(x) P(h)} \prod_{ij} \theta_{g_ijh_{ij}} \phi_{ijh_{ij}} \nonumber\\
  &= {1\over P(x) P(h)} \Bigl[ \prod_{rjt} \theta_{rjt}^{m_{rjt}} \Bigr]
     \Bigl[ \prod_{ij} \phi_{ijh_{ij}} \Bigr],
\end{align}
where we have used Eq.~\eqref{eq:likelihood} in the second line and once again assumed a uniform prior on~$\theta$.  This allows us, for example, to compute a MAP estimate of~$\theta_{rjt}$ by differentiating while imposing the constraint $\sum_t \theta_{rjt} = 1$ with a Lagrange multiplier, which yields the intuitive expression
\begin{equation}
\hat{\theta}_{rjt} = {m_{rjt}\over n_r}.
\end{equation}

Alternatively, if we have only the component assignments~$g$---from a consensus clustering for example---but no slot assignments, then we can write
\begin{align}
P(\theta|x,g) &= {P(x|g,\theta) P(\theta)\over P(x)}
   = {P(\theta)\over P(x)} \sum_{\set{h_{ij}}} P(x,h|g,\theta) \nonumber\\
  &= \sum_{\set{h_{ij}}} \prod_{ij} \theta_{g_ijh_{ij}} \phi_{ijh_{ij}}
   = \prod_{ij} \sum_{t=0}^{T_j-1} \theta_{g_ijt} \phi_{ijt} \nonumber\\
  &= \prod_{rj} P(\theta_{rj}|x,g),
\end{align}
with
\begin{equation}
P(\theta_{rj}|x,g) = \prod_{i\in r} \sum_{t=0}^{T_j-1} \theta_{g_ijt} \phi_{ijt},
\label{eq:ptheta}
\end{equation}
where the product is over observations~$i$ assigned to component~$r$, and we have again neglected multiplicative constants.  The separation of variables by component/item combinations $r,j$ makes finding a MAP estimate relatively straightforward, by maximizing~\eqref{eq:ptheta}.  The function is concave everywhere in the domain of~$\theta_{rjt}$ and readily (twice) differentiable, so many standard convex optimization methods can be applied.  One simple approach is to again use Jensen's inequality, Eq.~\eqref{eq:jensen}, and write
\begin{align}
\log P(\theta_{rj}|x,g) &= \sum_{i\in r} \log \sum_t \theta_{rjt} \phi_{ijt}
  \nonumber\\
  &\ge \sum_{i\in r} \sum_{t=0}^{T_j-1} q^{ij}_t \log {\theta_{rjt} \phi_{ijt}
       \over q^{ij}_t},
\label{eq:thetajensen}
\end{align}
where $\sum_t q^{ij}_t = 1$ and the exact equality is achieved when
\begin{equation}
q^{ij}_t = {\theta_{rjt} \phi_{ijt}\over\sum_{u=0}^{T_j-1} \theta_{rju} \phi_{iju}}.
\label{eq:qt}
\end{equation}
Following the same logic as in Section~\ref{sec:em}, we can then maximize~\eqref{eq:ptheta} by separately and repeatedly maximizing the right-hand side of~\eqref{eq:thetajensen} with respect to~$q^{ij}_t$, which we do using Eq.~\eqref{eq:qt}, and with respect to~$\theta_{rjt}$, which we do by differentiating.  Enforcing the constraint $\sum_t \theta_{rjt} = 1$ with a Lagrange multiplier, this gives
\begin{equation}
\theta_{rjt} = {1\over n_r} \sum_{i\in r} q^{ij}_t.
\label{eq:thetaem}
\end{equation}
Starting from any convenient initial values of the $\theta_{rjt}$, such as $\theta_{rjt} = 1/T_j$ for all $r,j,t$, applying Eqs.~\eqref{eq:qt} and~\eqref{eq:thetaem} in turn and iterating, we are guaranteed to converge to the global optimum because of the concavity of the objective, and hence we get our MAP estimate for~$\theta_{rjt}$.  Once we have estimates of the parameters, the functional forms of the probability distributions for the data can be reconstructed from Eq.~\eqref{eq:pxrj}.

We can also estimate the number of components from the output of the Monte Carlo algorithm.  Each sampled configuration $(k,g,h)$ carries its own value of~$k$, drawn from the distribution of Eq.~\eqref{eq:posterior}.  This allows us to trivially estimate the posterior distribution of~$k$ itself by constructing a histogram of the sampled values, or to make a MAP estimate of~$k$, which is simply the value that occurs most often among all samples.  Thus the Monte Carlo allows one both to assign observations to components and to perform model selection over the number of components in a single run of the algorithm.  No additional model selection step, for instance using the Bayesian information criterion, is necessary.

\begin{figure*}
\begin{center}
\includegraphics[width=8.6cm]{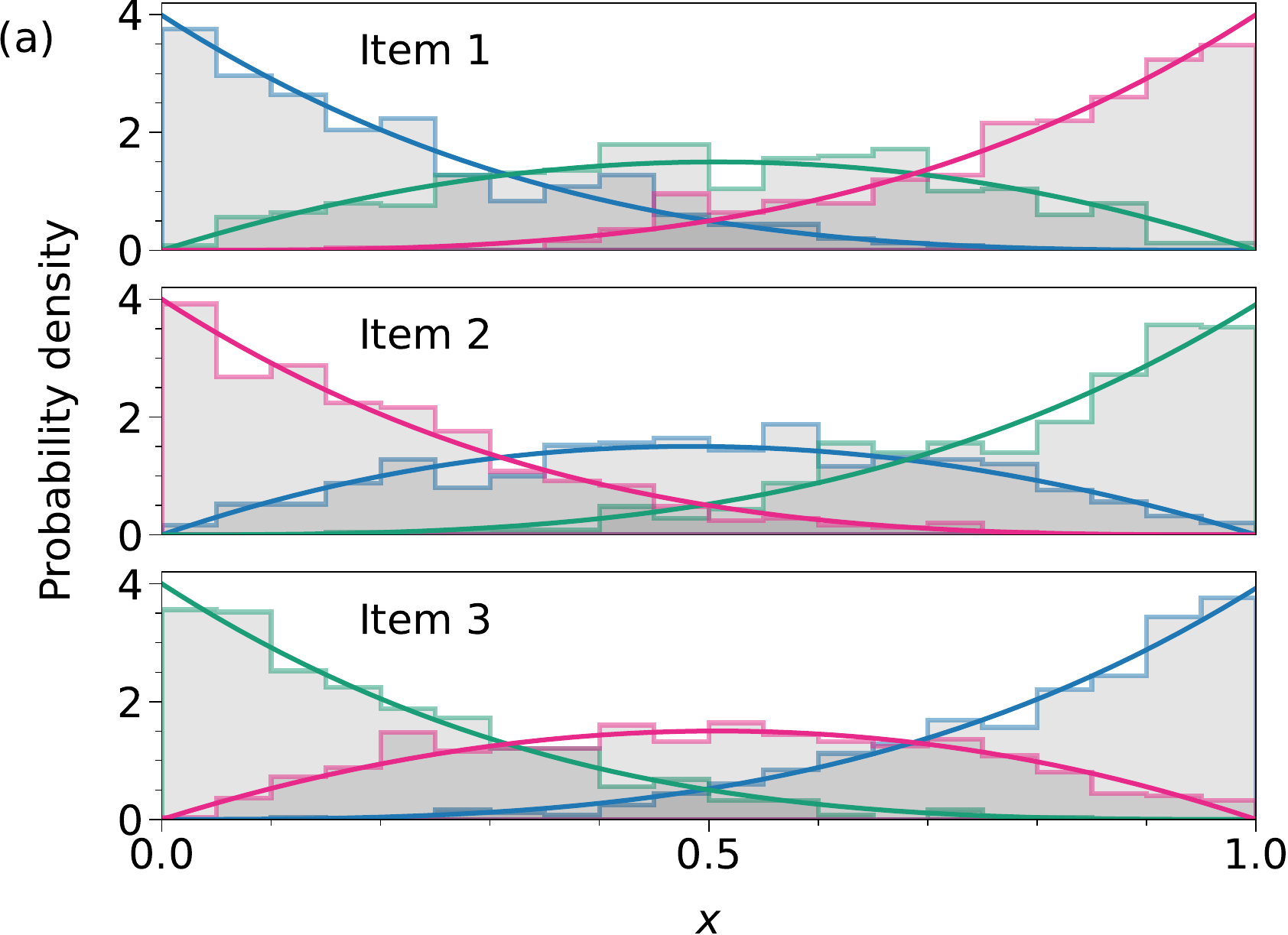}\hfill
\includegraphics[width=8.6cm]{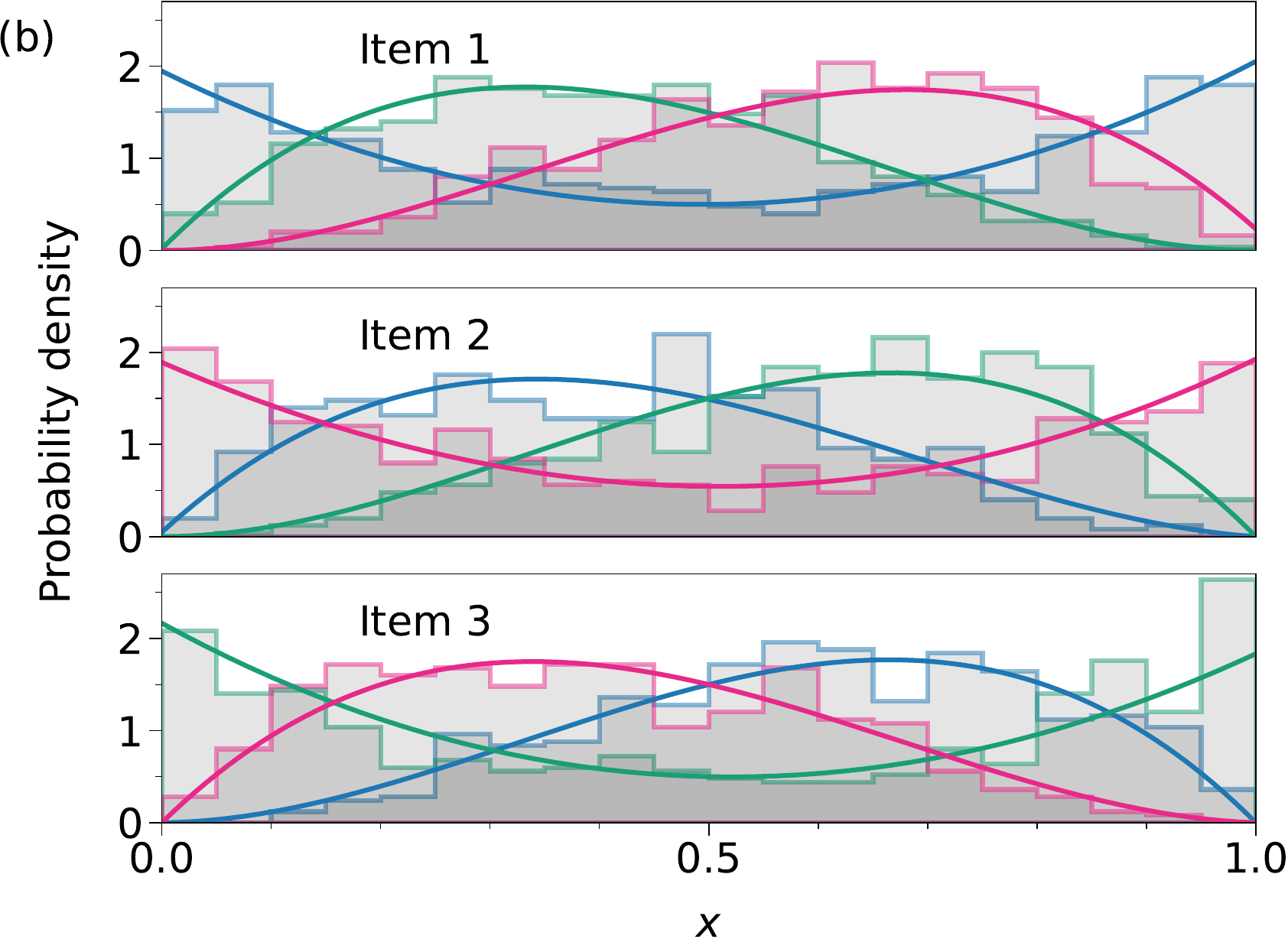}
\end{center}
\caption{(a)~The distributions of the observations in the synthetic data set of Eq.~\eqref{eq:synth1}, calculated using the EM algorithm of Section~\ref{sec:em}.  The histograms represent the actual distributions of the data; the curves represent the polynomial probability densities inferred by the algorithm.  Each panel represents one item and the colors indicate the three components.  (b)~Distributions for the data set of Eq.~\eqref{eq:synth2}.}
\label{fig:emsynth}
\end{figure*}

\section{Results}
\label{sec:results}
In this section we give example applications of our methods, illustrating their ability to capture hidden structure, infer the number of components and the data distribution for each component, and to accurately fit non-Gaussian data.  In order to perform these calculations we must make some choice for the basis functions.  In our examples we use the beta-function Bernstein basis of Eq.~\eqref{eq:bernstein} for data with bounded domain and an analogous basis of gamma functions for unbounded data.  We must also choose a prior~$P(k)$ on the number of groups and for this we use the uniform prior of Eq.~\eqref{eq:uniform} in all cases.

\subsection{Synthetic data}
\label{sec:synthetic}
For our first example, we perform tests on a selection of synthetic data sets, which provide a controlled laboratory for probing the performance of our methods.  The data in this section are themselves generated from the model of Section~\ref{sec:model} with Bernstein basis functions, and we test the ability of our algorithms to correctly recover the structure planted in the data, so these examples function as consistency tests of our approach.

As a first example, consider Fig.~\ref{fig:emsynth}.  For this test, we generated a data set with $N=1500$ observations of $M=3$ items each, divided in $k=3$ groups of 500.  Observations in the first group were drawn from three cubic probability distributions for the three different items:
\begin{subequations}
\label{eq:synth1}
\begin{align}
P_1(x) &= \Phi_0(x), \\
P_2(x) &= \tfrac12 \bigl[ \Phi_1(x) + \Phi_2(x) \bigr], \\
P_3(x) &= \Phi_3(x),
\end{align}
\end{subequations}
with $\Phi_t(x)$ being a Bernstein basis function as in Eq.~\eqref{eq:bernstein}, with degree $d=3$ (equivalent to setting $T_j=4$ in our general notation).  For the other two groups, observations are drawn from cyclic permutations of the same three functions.

Figure~\ref{fig:emsynth}a shows results from an analysis of these data using the EM algorithm of Section~\ref{sec:em}.  The three panels show the distribution of observations in the three components: the histograms show the actual distribution of the data in the ground-truth components while the curves show the distributions inferred by the algorithm, reconstructed from the fitted values of the parameters~$\theta_{rjt}$ using Eq.~\eqref{eq:pxrj}.  As the figure shows, the algorithm has ably discovered the components and their distributions.  Bear in mind that the algorithm does not know \textit{a~priori} which observations belong to which components, so it cannot calculate the histograms we see in the figure: it receives only unlabeled data and has the task of simultaneously dividing them into components and estimating the distributions.  If we assign each observation to the component with the highest estimated probability in the EM algorithm, the algorithm puts 1418 of the 1500 observations in the correct group, or 94.5\%.

Now consider Fig.~\ref{fig:emsynth}b, which shows results for a second data set of the same size and structure, but with the data in the first group drawn from the distributions
\begin{subequations}
\label{eq:synth2}
\begin{align}
P_1(x) &= \tfrac12 \bigl[ \Phi_0(x) + \Phi_3(x) \bigr], \\
P_2(x) &= \Phi_1(x), \\
P_3(x) &= \Phi_2(x),
\end{align}
\end{subequations}
and those in the other two groups having cyclic permutations of the same three distributions.  As the figure shows, the algorithm has again successfully determined the distributions for each component, but an important difference is that one distribution in each component is now bimodal, placing most weight at the extremes $x=0,1$ and relatively little in the middle of the range.  This is an example of a data set that could not be well fitted with a Gaussian mixture model, which only admits unimodal data.  A~Gaussian model would require at least four components to fit these data, and conventional model selection criteria would presumably (and incorrectly) favor a four-component model over the three-component one.

The EM algorithm is fast enough for convenient interactive use.  The runs that produced Fig.~\ref{fig:emsynth} took about 1~second each to calculate the probability parameters to an accuracy of~$10^{-10}$ on the author's (up-to-date but otherwise unremarkable) laptop computer.

\begin{figure}
\begin{center}
\includegraphics[width=8.6cm]{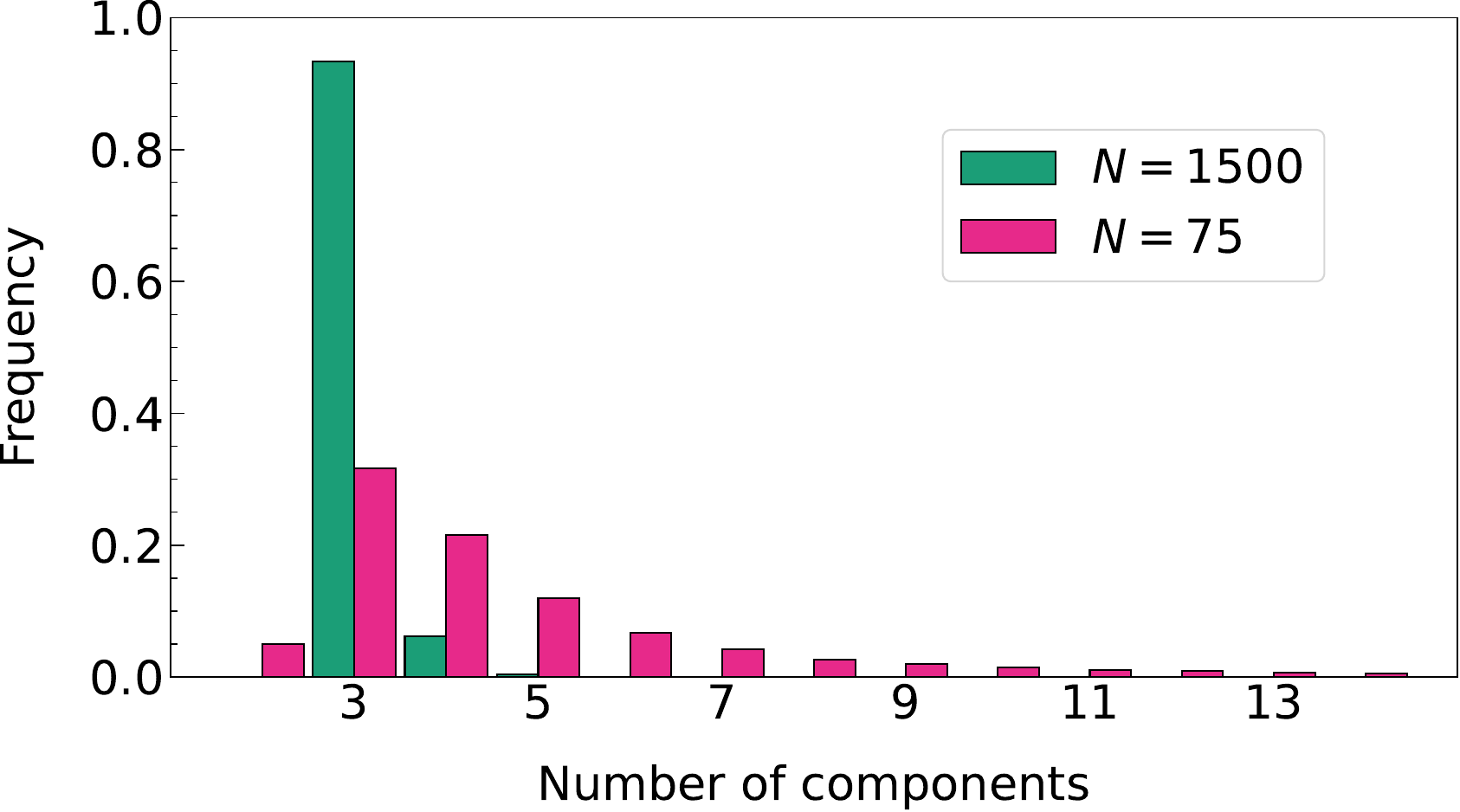}
\end{center}
\caption{The distribution of the number of components~$k$ in Monte Carlo samples for the same synthetic data set as in Fig.~\ref{fig:emsynth}a with $N=1500$, and for a smaller data set with the same parameters but $N=75$.  The true number of components in both cases is three.}
\label{fig:mcsynth}
\end{figure}

Turning to the Monte Carlo algorithm, a particular feature of that algorithm is its ability to infer not only the assignment of observations to components but also the number of components.  Figure~\ref{fig:mcsynth} shows results from a single run of the Monte Carlo algorithm on the same data set as Fig.~\ref{fig:emsynth}a, with 2500 sweeps for burn-in followed by 25\,000 for sampling (green bars in the figure).  The plot is a histogram of the sampled values of the number of components~$k$ and, as we can see, the algorithm has correctly inferred that there are three components in this data set: $k=3$ is strongly favored over any other value.

The algorithm's ability to do this does depend on the quality of the data.  For instance, the second histogram in Fig.~\ref{fig:mcsynth} (in magenta) shows results for a data set with the same parameters but only 75 observations in three groups of 25 each.  As we can see, $k=3$ is still the most likely number of components but there is much more uncertainty, with values from 2 up to 14 a possibility.

The Monte Carlo algorithm is more computationally demanding than the EM algorithm but still fast enough for interactive use.  For the larger ($N=1500$) data set above, our calculation took about 5~seconds on the author's laptop.  For data sets of this size the decision of whether to use the EM or Monte Carlo algorithm will hinge more on the type of results one wants than on running time.

\subsection{Real-world data}
\label{sec:realdata}
In this section we give a number of example applications to real-world data from various domains.

\subsubsection{Italian wine}
\label{sec:wine}
Our first application is to a classic data set, frequently studied in the machine learning literature, consisting of measurements for 178 wines of 13 real-valued chemical and photometric properties, including alcohol content, malic acid content, phenol content, color intensity, and hue.  All the wines were grown in the same region of Italy, but they were made from three different grape varietals, and previous analyses have shown that it is possible to infer the grapes from the measured quantities.  Further details of the data set are given in Appendix~\ref{app:data}.

\begin{figure}
\begin{center}
\includegraphics[width=8cm]{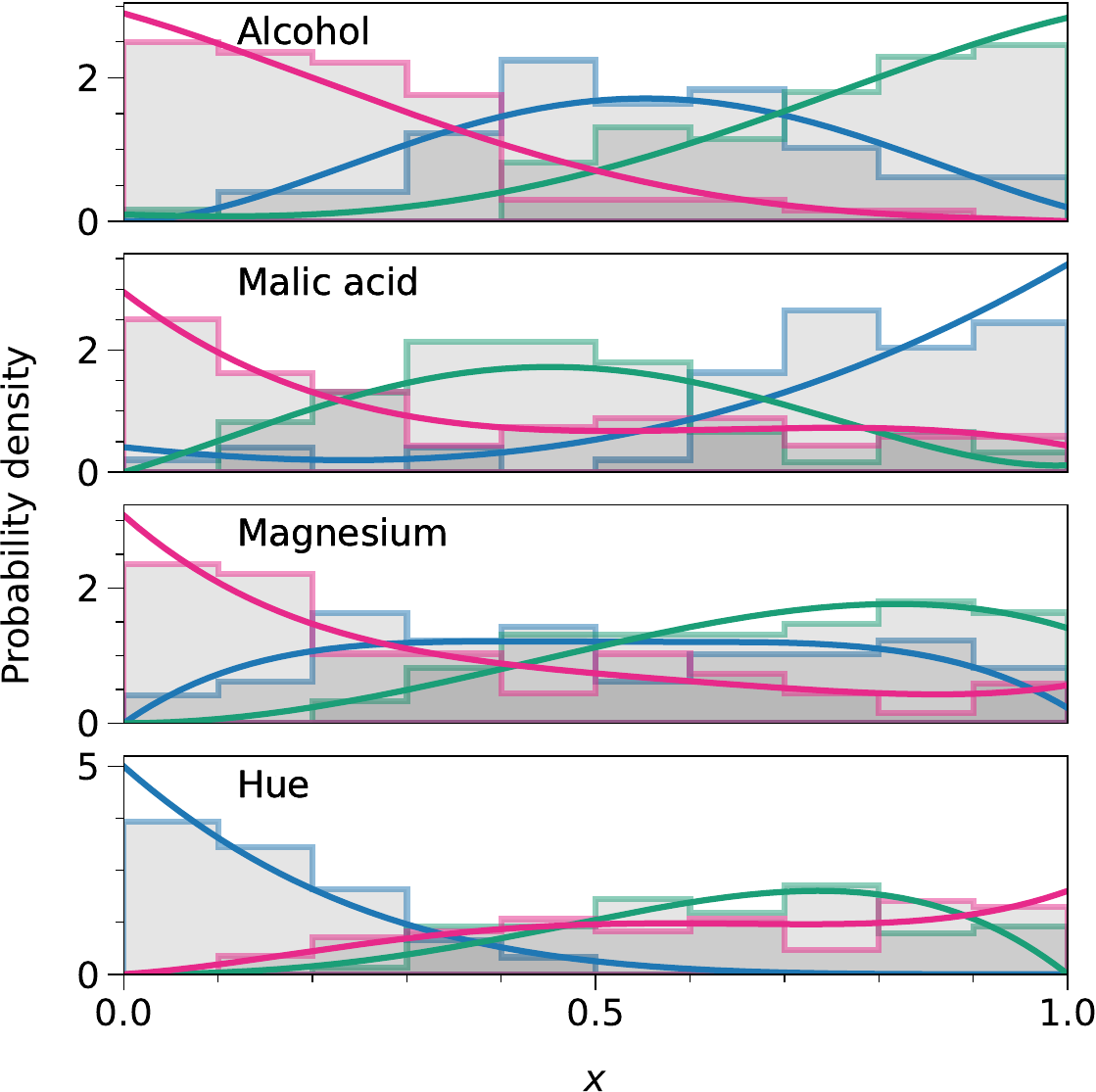}
\end{center}
\caption{Actual (histograms) and inferred (curves) distributions of four of the measures in the wine data set, for each of the three inferred components.}
\label{fig:wine}
\end{figure}

Before applying our methods to these data, we first map each of the 13 measures individually to the interval~$[0,1]$.  There are many ways of doing this.  The simplest is a linear rescaling that places the lowest value at 0 and the highest at~1.  This is not always ideal, however.  For data sets that contain occasional outliers, as this one does, such a rescaling can place most of the data in a relatively narrow numeric range and obscure the interesting details of their structure.  A simple and effective alternative is to use the cumulative distribution function (CDF) of the data instead of the raw data themselves, which has the dual advantages of automatically mapping the data to $[0,1]$ and also making them uniform in that interval.  (A potential disadvantage, worth keeping in mind, is that it also disguises outliers, which might be important in some cases.)

Performing this transformation for each of the 13 wine measures, we then apply our EM algorithm with $k=3$ components and Bernstein basis functions with degree~$d=4$ for all variables (equivalent to $T_j=5$).  Running time is about 3~seconds and the result is a very clear separation of the wines into the three components, with most wines being assigned with over 99\% confidence to a specific component.  Comparing these assignments with the ground-truth data on the grape varietals, we find that 175 of the 178 wines are classified into the correct group.  The results are not highly sensitive to the polynomial degree, although we do find slightly poorer classification performance for $d=3$ (172 out of 178 wines correct).  Figure~\ref{fig:wine} shows a selection of the inferred probability distributions calculated by the method, along with histograms of the actual data.  For some quantities, such as alcohol content, the distributions reflect a simple low/medium/high division among components, but for others, such as magnesium content and hue, they are more varied.  Our method is particularly useful for highlighting distinctions such as this.  Of particular note is the fact that many of the distributions are clearly non-Gaussian, and would not be well fit by a conventional Gaussian profile analysis.

\begin{figure}
\begin{center}
\includegraphics[width=\columnwidth]{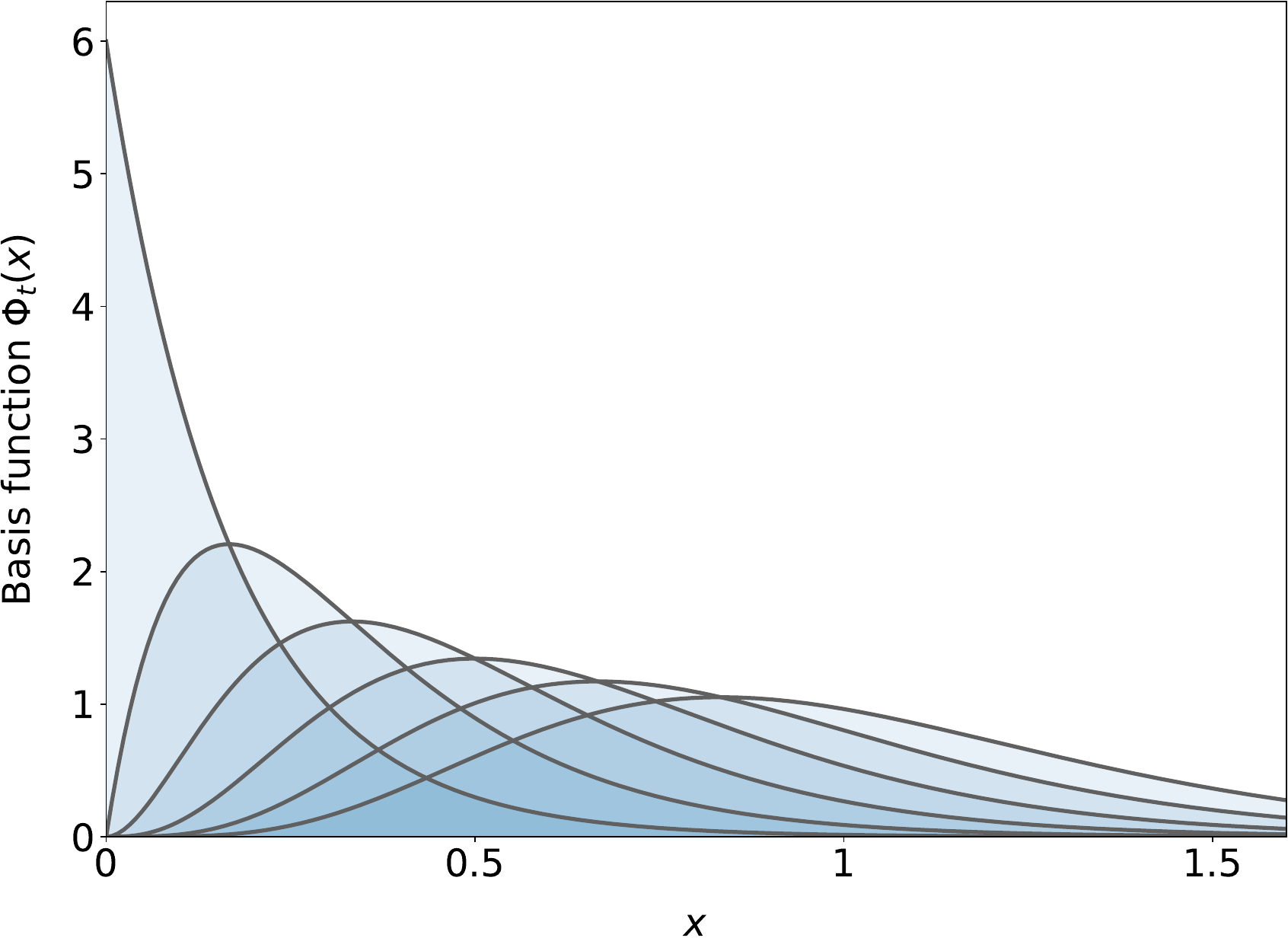}
\end{center}
\caption{The gamma-function basis of Eq.~\eqref{eq:gammas} for $T=5$.}
\label{fig:gammas}
\end{figure}

\subsubsection{Cancer biomarkers}
\label{sec:biomarkers}
Our second example is a clinical one and illustrates the use of a different basis set.  The data are from Debernardi~\etal~\cite{Debernardi20} and consist of measurements of four diagnostic biomarkers for early-stage pancreatic cancer (creatinine, LYVE1, REG1B, and TFF1) in 590 patients, including individuals who have and have not been diagnosed with cancer.  Debernardi~\etal\ provide a ``diagnosis'' variable for each patient, which has three values: no pancreatic disease, benign disease, and cancer.  Details of the data set are given in Appendix~\ref{app:data}.

All the measured values are non-negative and all are concentrated around zero, but with tails of higher values.  For data of this kind we employ a basis set of gamma functions:
\begin{equation}
\Phi_t(x) = {(xT)^t\over t!}\,T\e^{-xT} \qquad\mbox{for $t = 0\ldots T-1$.}
\label{eq:gammas}
\end{equation}
Like the Bernstein basis, these functions are single-peaked with equally spaced maxima in $[0,1]$, but unlike the Bernstein basis they are unbounded above, having an exponential tail.  They form a complete basis for functions of the form $f(x)\,\e^{-xT}$ where $f(x)$ is a polynomial of degree~$T-1$, and are suitable for representing non-negative data that fall primarily in the range~$[0,1]$ but may extend arbitrarily far above that range on occasion.  A plot of the functions for $T=5$ is shown in Fig.~\ref{fig:gammas}.

\begin{figure}
\begin{center}
\includegraphics[width=8cm]{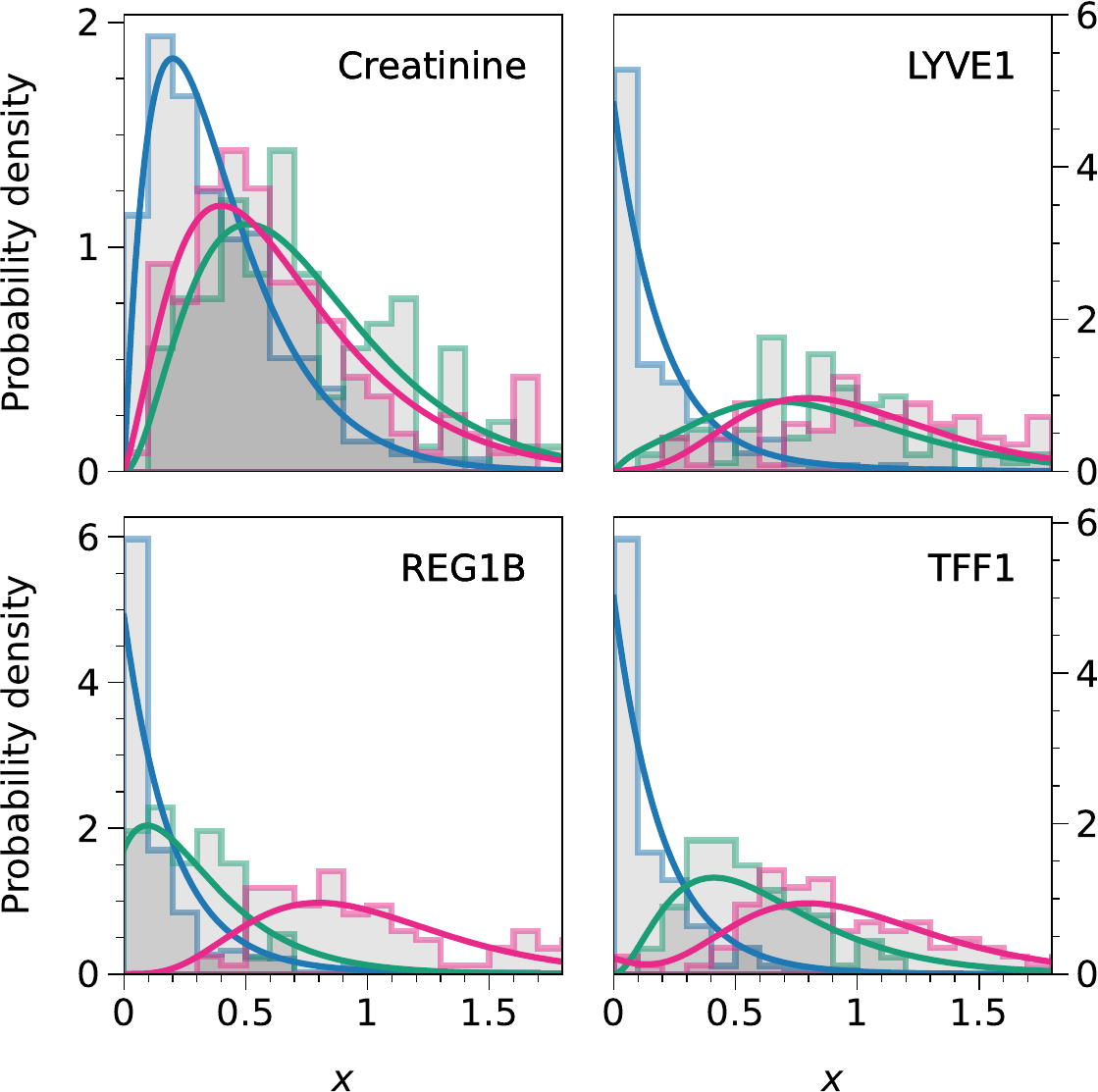}
\end{center}
\caption{Actual (histograms) and inferred (curves) distributions of the four biomarkers used in the pancreatic cancer data set, for each of the three inferred components.}
\label{fig:empancr}
\end{figure}

The raw biomarker data have various ranges, but for our analysis we rescale each variable so that its mean is~$\frac12$.    We run our EM algorithm on the data with $k=3$ using the gamma function basis above with $T_j=5$ for all variables~$j$.  (As mentioned in Section~\ref{sec:model}, the code for the algorithm works unmodified with any set of basis functions---we only need to calculate the values of the quantities~$\phi_{ijt}$, Eq.~\eqref{eq:phi}.)  The calculation takes about 10 seconds and produces the results shown in Fig.~\ref{fig:empancr}.

As the figure shows, the analysis finds a clear separation between a group of patients for whom all biomarker values are low (blue curves), and the remainder of the population, who are divided into two groups (green and magenta) with higher values that are less well separated, being mainly distinguished by their REG1B values and, to a lesser extent, TFF1.  Assuming our goal is to identify potential cancer cases, and if we tentatively identify the latter two groups as suspect and the first group as not, then the algorithm has relatively good classification performance for an unsupervised method.  It places 309 out of 391 non-cancer patients in group~1 and 131 out of 199 cancer patients in groups 2 and~3, giving a sensitivity of 66\% and specificity of~79\%.

\begin{figure}
\begin{center}
\includegraphics[width=8.3cm]{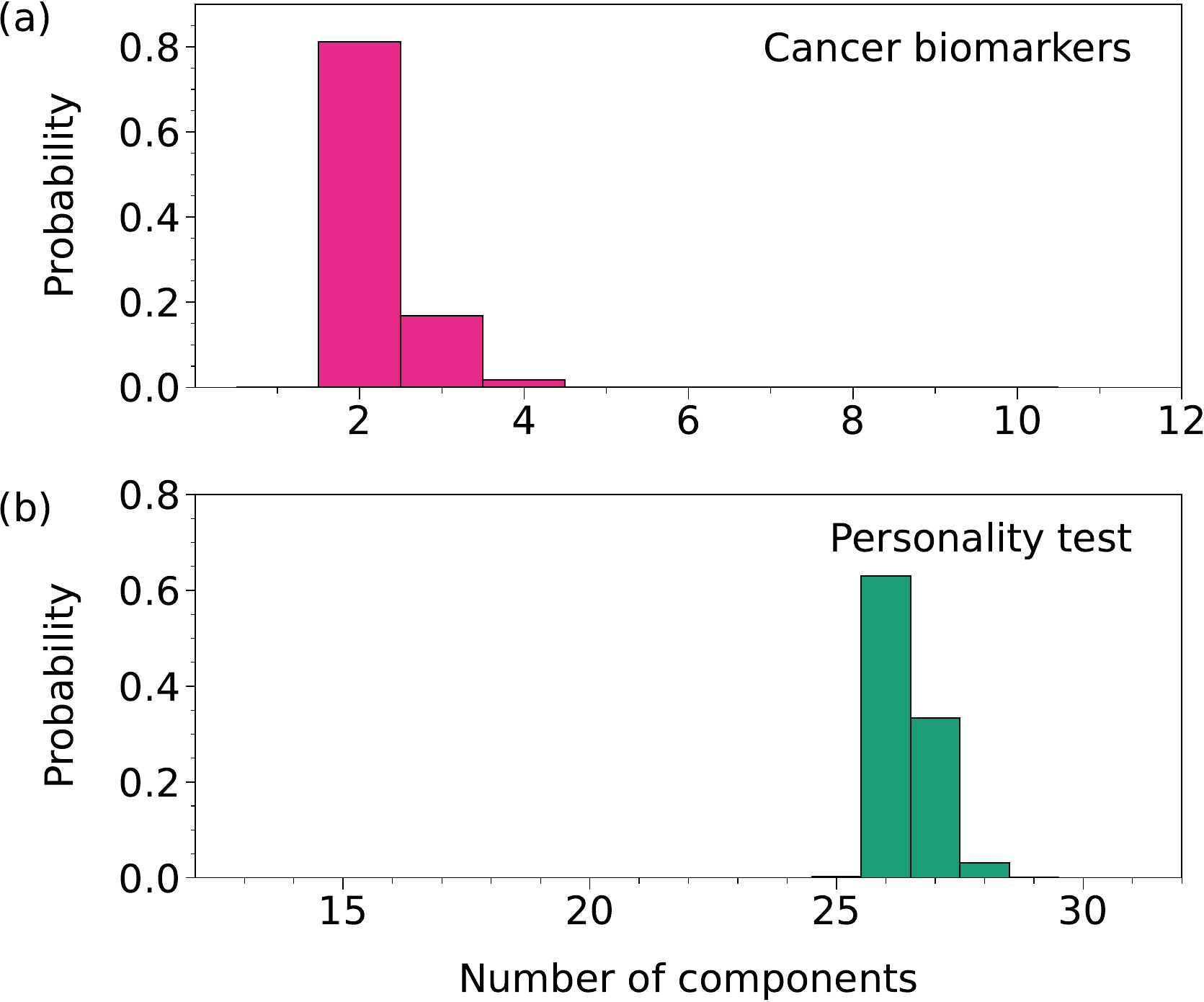}
\end{center}
\caption{Inferred number of components in (a)~the biomarkers data set and (b)~the personality test data set.}
\label{fig:khist}
\end{figure}

Given the similarity of the profiles for groups 2 and~3, one could reasonably ask whether the data actually support three groups in this case.  This question can be addressed using our Bayesian Monte Carlo analysis.  Applying our Monte Carlo algorithm, again with $T_j=5$ for all variables, we run for 5000 sweeps of burn-in followed by 50\,000 sweeps for sampling.  The calculation takes about 5~seconds.  Figure~\ref{fig:khist}a shows the distribution of sampled values of the number of components~$k$ and, as we can see, the most probable number of components is in fact $k=2$, although $k=3$ carries some weight also and is not ruled out.

\begin{figure}
\begin{center}
\includegraphics[width=\columnwidth]{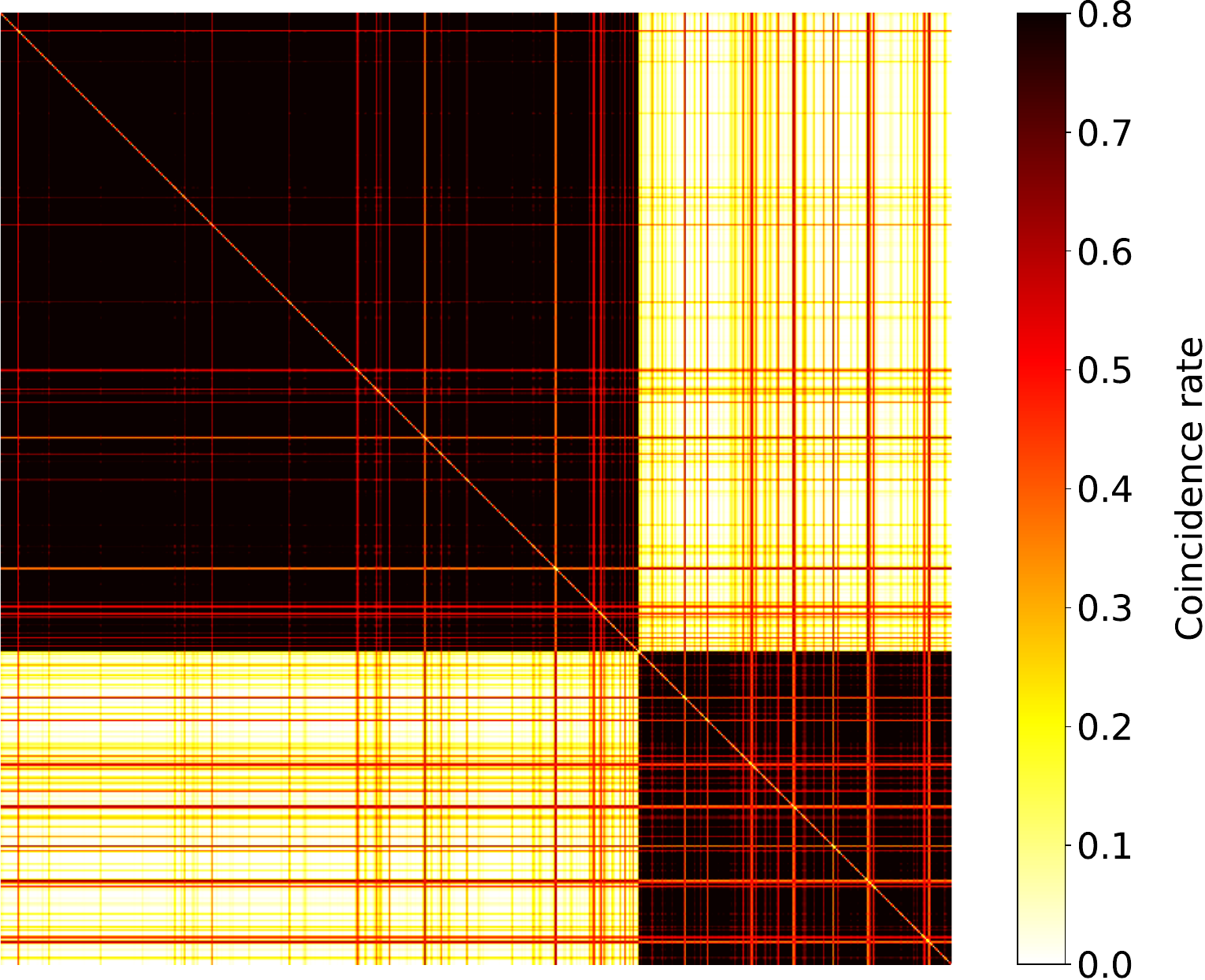}
\end{center}
\caption{The consensus matrix for the biomarkers data set, with matrix values (which lie between zero and one) represented using the color code on the right.  Values in this matrix indicate the fraction of samples in which a given pair of individuals are placed in the same component.  The rows and columns of the matrix have been permuted to place the individuals in each component next to one another, and the components then appear as the darker blocks.}
\label{fig:panmatrix}
\end{figure}

Given these findings, one might reasonably now rerun the EM analysis with $k=2$ instead of $k=3$, but for illustrative purposes let us instead use the output of the Monte Carlo to group the patients by constructing a consensus clustering based on the sampled configurations.  Of the many ways to do this, we employ one of the simplest and most effective~\cite{DN07,ZDA23}, in which we first compute a consensus matrix, which is the $N\times N$ symmetric matrix with elements~$C_{ij}$ equal to the fraction of samples in which individuals $i$ and~$j$ are assigned to the same component.  We construct such a matrix for the complete set of samples and also for each individual sample, then find the sample whose individual matrix has the smallest mean-square distance from the matrix of the complete set.  Figure~\ref{fig:panmatrix} shows a graphical rendering of the consensus matrix in which the rows and columns have been arranged to group the members of each consensus component together, and the components emerge clearly as the two prominent blocks.  In this case the two-component structure of the data is very clear.  Comparing the consensus clustering to the ground-truth data on diagnoses, we find a classification performance similar to the EM algorithm, with a slightly lower sensitivity of 63\% but a slightly higher specificity of 83\%.  Classification is not our main goal here---we are interested in our ability to fit mixtures of arbitrary data distributions, and in any case better performance could surely be achieved using supervised learning---but these results suggest that the method could be a useful tool for unsupervised problems or exploratory analysis.

\subsubsection{Personality test}
\label{sec:big5}
For our last example we illustrate the application of our methods to a substantially larger data set.  The data come from an online personality test and describe the responses of 10\,000 participants when asked whether they agree with 50 statements about themselves, such as ``I am the life of the party,'' or ``I am always prepared.''  Responses are on a scale of 1 to~5, with 1 meaning strongly disagree and 5 meaning strongly agree.  We map these responses to the interval $[0,1]$ with values 0.1, 0.3, 0.5, 0.7, and 0.9.  The original test had the goal of measuring where each respondent fell with respect to the ``Big~5'' personality traits of openness, conscientiousness, extroversion, agreeableness, and neuroticism~\cite{Digman90}, and the questions were designed with this goal in mind.  The data set is described in more detail in Appendix~\ref{app:data}.

We do not know \textit{a~priori} how many components there should be in this data set so we take a Bayesian approach.  Using Bernstein basis functions with degree $d=2$, we run our Monte Carlo algorithm for 2500 sweeps of burn-in followed by 25\,000 sweeps for sampling.  We choose $d=2$ because there are only five distinct values of the observed variables and we want the number of basis functions to be less than this to avoid overfitting.  The calculation takes considerably longer than for the previous data sets, partly because a sweep now corresponds to 10\,000 Monte Carlo steps and partly because the number of survey items and number of components are both large, which increases the computational effort per step.  The entire run took about 18~minutes on the author's laptop.

Figure~\ref{fig:khist}b shows the resulting distribution of the number of classes and the calculation favors about 26 components for this data set, a much larger number than in our other examples, indicating perhaps that there are 26 distinguishable personality types in the data.  If the Big~5 theory were correct, and if every combination of the five traits were possible, there would be $2^5=32$ different personality types, which is quite close to the observed number, although this is probably just coincidence.

A problem of interest that we can tackle using the Bayesian approach is \textit{variable selection}, the question of which of the observed variables are most informative about component assignments and hence most useful for classifying observations.  In the present case, for example, if one could rule out some variables as being mostly uninformative, one could simplify future tests to focus on the informative variables.  There are various ways to perform variable selection but a simple one in the present case is to look at the mutual information between the slots and the component assignments.  The slot assignments encode the data, since they tell us from which basis function the data were drawn, and mutual information measures precisely how much one random variable tells us about another.  The general expression for the mutual information between two random variables~$x$ and~$y$ is~\cite{CT06}
\begin{equation}
I(x;y) = \sum_{xy} P(x,y) \log {P(x,y)\over P(x) P(y)},
\end{equation}
where $P(x)$, $P(y)$, and $P(x,y)$ are the marginal and joint distributions of the variables.  In our case $x$ is a component assignment~$r$ and $y$ is a slot~$t$.  The probability distributions for item~$j$ are
\begin{align}
P_j(r,t) &= {1\over N} \sum_i \delta_{g_ir} \delta_{h_{ij}t} = {m_{rjt}\over N}, \\
P_j(r) &= {1\over N} \sum_i \delta_{g_ir} = {n_r\over N}, \\
P_j(t) &= {1\over N} \sum_i \delta_{h_{ij}t} = {n_{jt}\over N},
\end{align}
where $n_{jt}$ is the number of individuals (in any component) assigned to slot~$t$ for item~$j$.  Then the mutual information for item~$j$ is
\begin{equation}
I_j(r;t) = {1\over N} \sum_{rt} m_{rjt} \log {Nm_{rjt}\over n_r n_{jt}},
\end{equation}
and we average this expression over all samples to get the average mutual information.  High values of the resulting quantity indicate variables that are highly informative about which components observations belong to.

\begin{figure}
\begin{center}
\includegraphics[width=\columnwidth]{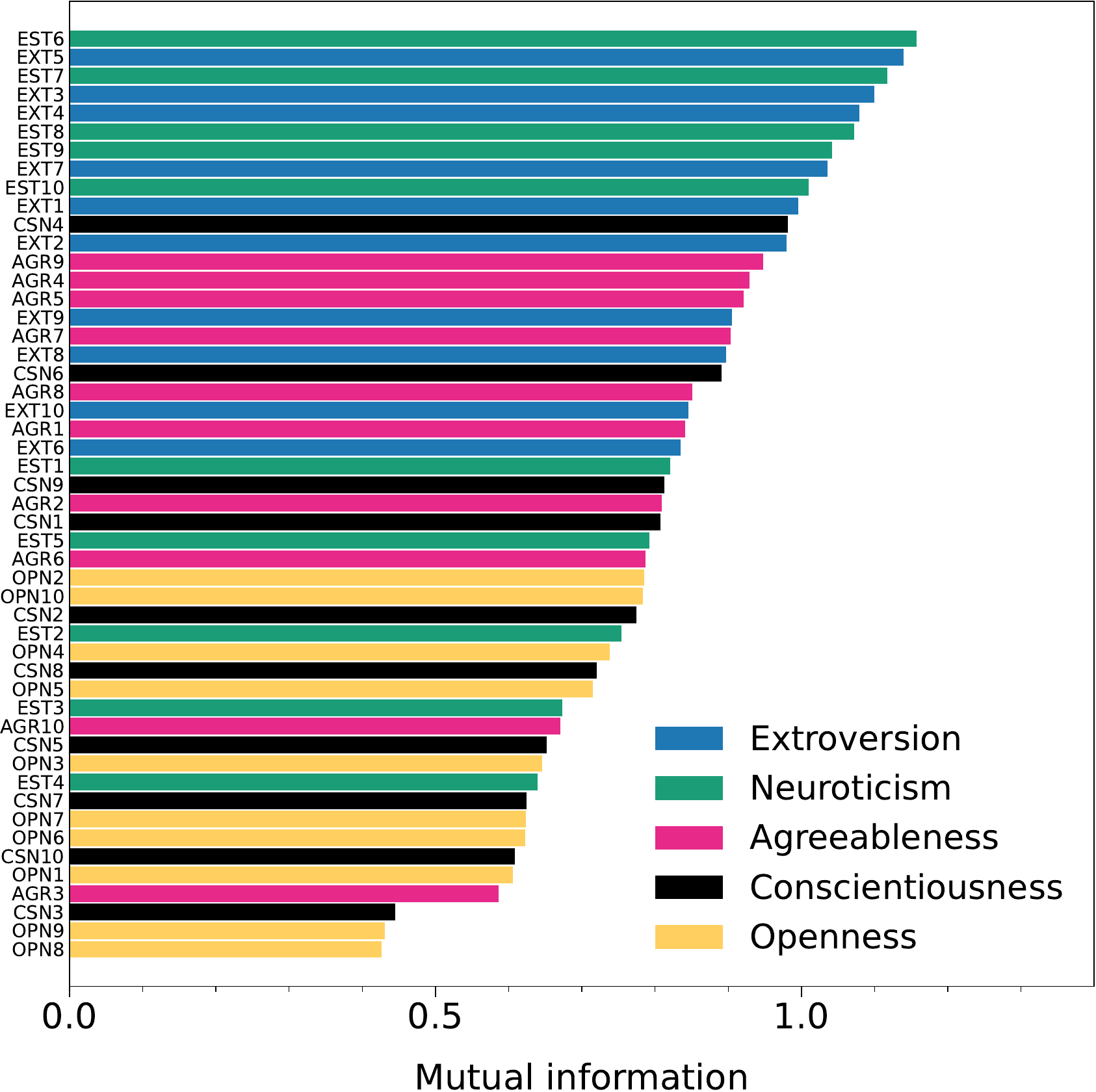}
\end{center}
\caption{Average mutual information in bits between component and slot assignments for the 50 items in the personality test data set.  The mutual information is a measure of how informative the data are about which component a respondent is assigned to.  The wording of the test questions corresponding to each of the item codes can be found in Table~\ref{tab:big5} of Appendix~\ref{app:data}.}
\label{fig:big5vars}
\end{figure}

Figure~\ref{fig:big5vars} shows the mutual information between components and slots for each of the 50 variables.  Each question in the test was designed to probe one of the Big~5 traits and the bars in the figure are color coded by intended trait.  As the figure shows, the top half of the plot---the most informative variables by this measure---are primarily those associated with extroversion, neuroticism, and, to a lesser extent, agreeableness, while most of the questions concerning conscientiousness and openness fall in the bottom half of the plot.  This observation could be taken as evidence confirming the validity of the extroversion, neuroticism, and agreeableness axes, while calling into question the conscientiousness and openness ones.  An alternative interpretation is that the trends we are seeing are associated with question order.  The experimenters report that they asked the questions in banks of five, with each bank containing one question probing each trait (see Table~\ref{tab:big5} in Appendix~\ref{app:data}), and always in the order extroversion-neuroticism-agreeableness-conscientiousness-openness, precisely the ordering seen in Fig.~\ref{fig:big5vars}.  It could be that the lower ranked traits are lower not because their validity is questionable but because the test takers took questions at the end of each bank less seriously than those at the start, and hence that the quality of the data is poorer.

Even if this were the case, however, the analysis is still a useful one.  Apart from highlighting possible issues with the data, it also offers evidence that at least some of the questions, those at the top of the ranking in Fig.~\ref{fig:big5vars}, are informative about personality classification.  The top five questions, for instance, all concerned with extroversion and neuroticism, are: {\sc est6} ``I~get upset easily,'' {\sc ext5} ``I~start conversations,'' {\sc est7} ``I change my mood a lot,'' {\sc ext3} ``I feel comfortable around people,'' and {\sc ext4} ``I~keep in the background.''  Somewhat lower on the list, but still having relatively high mutual information, are questions probing conscientiousness and agreeableness, such as {\sc csn4} ``I make a mess of things'' and {\sc agr9} ``I feel others' emotions.''  None of the questions probing openness have high mutual information, although values are not so low as to indicate that these questions should be removed from the test entirely.

\section{Discussion}
In this paper we have introduced a class of mixture models in which the data in each component of the mixture are drawn from an arbitrary probability distribution expressed as a linear combination of non-negative basis functions.  The primary application for these models is to multivariate latent profile analysis, and we have described two algorithms for performing analyses by fitting these models, one an expectation-maximization (EM) algorithm and the other a fully Bayesian algorithm making use of a collapsed Gibbs sampler.  The EM algorithm is numerically efficient, but returns only point estimates of the probability densities and a posterior distribution on component assignments conditioned on those densities, and requires the number of components to be externally specified by the user.  The Bayesian approach, on the other hand, while more computationally intensive, returns a full (unconditional) posterior distribution on the component assignments, and can in addition be used to sample, and hence select, the number of components.  We have given a number of example applications of our methods, showing that we can recover known planted structure in synthetic tests and successfully classify data in real-world examples.

One potential criticism of our approach is that inferring a set of complex probability distributions for each component requires more statistical power than, for example, inferring a simple Gaussian, and is more susceptible to uncertainty when that power is lacking.  This is undoubtedly true.  Each probability distribution in our model is characterized by $T_j$ parameters~$\theta_{rjt}$, where a Gaussian requires only two parameters.  More data, or higher-quality data, will be required to fit the former model than the latter.  On the other hand, one can equally say that if we have sufficient statistical power to fit a Gaussian model but not a non-Gaussian one, then we also lack the power to rule out the Gaussian model if it is a poor fit, which limits its usefulness.

There are many additions and extensions of this work that would be possible.  Many standard measures of goodness of fit and diagnostic criteria can be straight\-forwardly applied to our models, including entropy, Bayes factors, approximate or numerical estimates of data evidence, and final data likelihood or posterior probability of the fitted model.  It would also be straightforward to generalize our approach to encompass data sets with mixed real-valued and categorical data, giving a hybrid of latent profile and latent class analysis.

One question whose answer appears less obvious is how to perform model selection on the number~$T_j$ of basis functions used to represent the profile probability distributions.  Larger numbers of basis functions give us the ability to capture more detail in these distributions, but the accompanying larger numbers of parameters may not be justified by the improvement in the fit.  In our work we have used fixed numbers of basis functions that we choose, but a more rigorous selection method would be desirable.  Previous work on choosing basis functions for density estimation could be a guide in this situation~\cite{WG19}, or a Bayesian sampling method might be possible, akin to our sampling of the number of components~$k$, but the workings of such a hypothetical method seem non-trivial.  We would expect an interaction between $T_j$ and~$k$ since larger values of~$T_j$ give us the ability to separate components whose distinction in the data relies on small-scale detail not visible in models with smaller~$T_j$.  So proper model selection on~$T_j$ would presumably require us to sample from the joint distribution of both $T_j$ and~$k$.  While intriguing, however, these questions we leave for future work.

\begin{acknowledgments}
The author thanks Max Jerdee and Alec Kirkley for useful conversations.  This work was supported in part by the US National Science Foundation under grant DMS--2404617.  Code implementing our methods is available at \verb|https://umich.edu/~mejn/lpa| for download.
\end{acknowledgments}

\appendix
\section{Basis functions}
\label{app:basis}
The theory developed in this paper is agnostic about the choice of basis functions for representing probability distributions.  One needs only to be able to evaluate the quantities $\phi_{ijt} = \Phi_{jt}(x_{ij})$ of Eq.~\eqref{eq:phi} and all other developments follow.  In the example applications of Section~\ref{sec:results} we use Bernstein and gamma bases, and these appear to work well for the particular data sets studied, but other choices may be appropriate in other circumstances.  Here we list a few of the many possibilities.

\subsection{Piecewise constant functions}
Perhaps the simplest choice of basis set is the set of ``top-hat'' functions
\begin{equation}
\Phi_t(x) = \Theta(x-t) \bigl[ 1 - \Theta(x-t-1) \bigr]
\label{eq:tophat}
\end{equation}
for $t = 0\ldots T$, where $\Theta(x)$ is the Heaviside step function, which is 1 if $x\ge0$ and 0 otherwise.  Then $\Phi_t(x)$ is 1 for $t\le x<t+1$ and zero elsewhere.  (We drop the subscript~$j$ on $\Phi_{jt}(x)$ in this appendix, since it plays no part in our discussions.)

Convex combinations of these functions produce piecewise constant probability densities spanning the range $x\in[0,T]$.  A mixture model using such a basis effectively assumes the density of observations in each unit interval to be independent of the density in all other intervals, with no constraints on continuity.  Alternatively, such a model is equivalent to a latent class model in which our real-valued observations are turned into categorical data by dividing them among a set of equal-width bins.  The particular definition in Eq.~\eqref{eq:tophat} assumes that the bins are of unit width, but it is a trivial matter to rescale either the basis functions or the data if some other bin width is desired.

\subsection{Bernstein basis polynomials}
For data on a bounded interval the Bernstein polynomials of Eq.~\eqref{eq:bernstein} and Fig.~\ref{fig:bernstein} are arguably a better choice than the top-hat functions of Eq.~\eqref{eq:tophat}.  Like the top hats, they are single-peaked localized functions, but they are smoother, providing a ``soft binning'' of the data and enforcing continuity on the probability density.  By controlling the number of basis functions, or equivalently the degree of the polynomials, one can control the effective number of bins and the degree of smoothness in the resulting functions.  The Bernstein basis has been used extensively for density estimation~\cite{Petrone99,Kakizawa04}, where it leads to a discrete form of beta-kernel estimation.

For certain types of data, special combinations of the Bernstein polynomials can be useful.  For data distributed symmetrically about the mid-point of the domain, perhaps because of physical or mathematical symmetries, combinations $\frac12[\Phi_t(x)+\Phi_{d-t}(x)]$ can form a useful basis.  For higher-dimensional or correlated data one can also form a multivariate Bernstein basis~\cite{WG19}.  A~bivariate probability density on $[0,1]^2$, for example, can be written~as
\begin{equation}
P(x,y) = \sum_{tu} \theta_{tu} \Phi_t(x) \Phi_u(y),
\end{equation}
with $\sum_{tu} \theta_{tu} = 1$.  The number of parameters~$\theta_{tu}$ can, however, grow rapidly as one extends this approach to higher degree or higher dimension, meaning that it is appropriate primarily for rich data sets that can justify such a large parameter space.

\subsection{Gamma functions}
For non-negative data on a semi-infinite interval a practical choice of basis set is the gamma distributions defined in Eq.~\eqref{eq:gammas} and depicted in Fig.~\ref{fig:gammas}.  Like the Bernstein basis, these functions are single-peaked with equally spaced maxima in $[0,1]$, but unlike the Bernstein basis they are unbounded above, having an exponential tail.  They are suitable for non-negative data that fall primarily in the range~$[0,1]$ but may extend arbitrarily far above that range on occasion.  Data with a different range can be easily rescaled to the appropriate size.  See Section~\ref{sec:realdata} for an example.

\begin{figure}
\begin{center}
\includegraphics[width=\columnwidth]{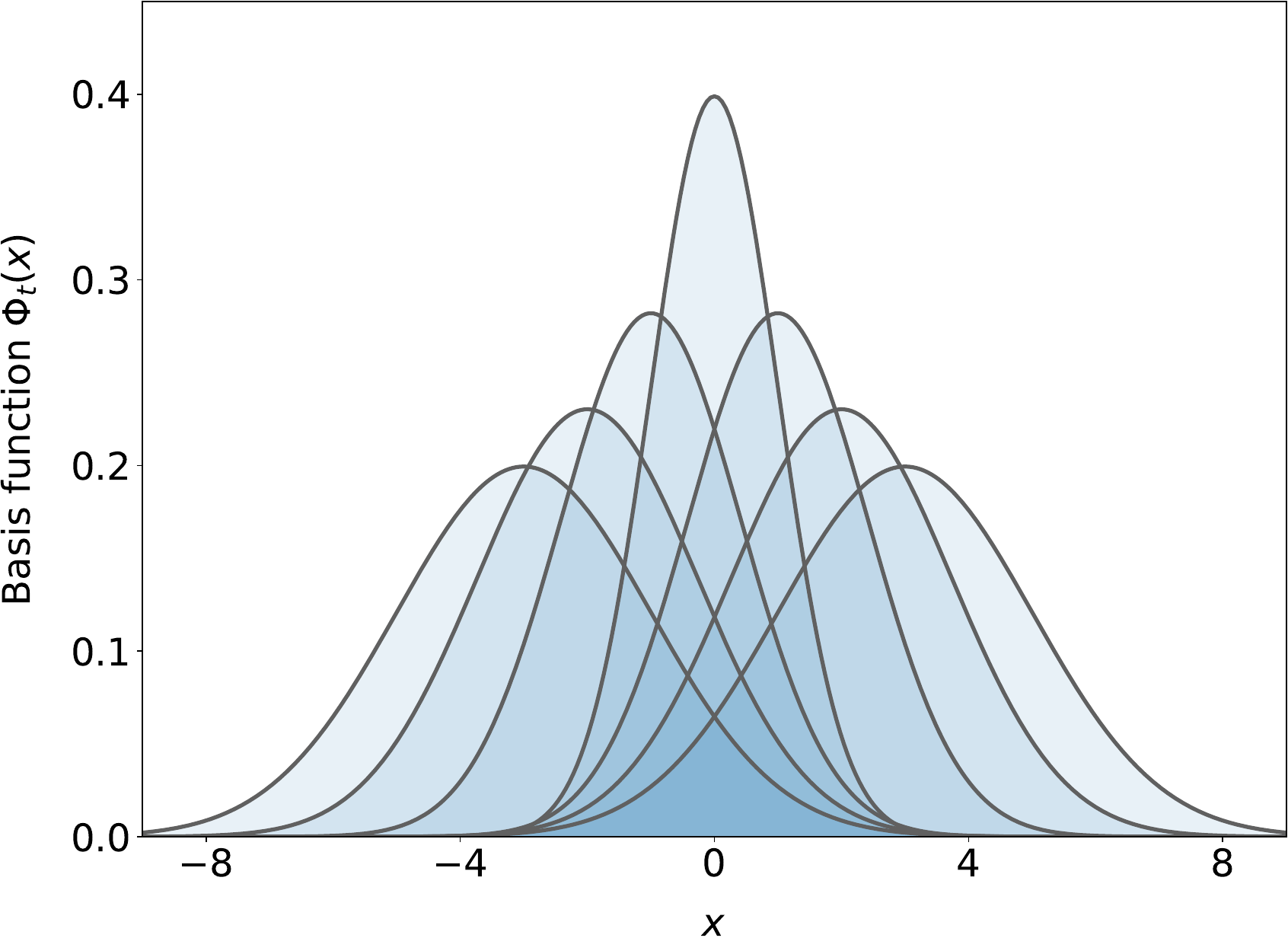}
\end{center}
\caption{A seven-member family of Gaussian basis functions, Eq.~\eqref{eq:gaussians}.}
\label{fig:gaussians}
\end{figure}

\subsection{Gaussians}
Gaussians have been used extensively in mixture-of-mixture models and profile analysis, but normally one fits the parameters of the Gaussians (means and variances) to the data.  One can use Gaussians with our approach too, but they then have fixed means and variances.  For instance, one possible choice of Gaussian basis would be
\begin{equation}
\Phi_t(x) = {1\over\sqrt{2\pi(|t|+1)}} \exp\biggl[ {-(x-t)^2\over2(|t|+1)}
             \biggr]
\label{eq:gaussians}
\end{equation}
for $t=-\frac12(T-1)\ldots\frac12(T-1)$, which is akin to a two-sided version of the gamma basis of Eq.~\eqref{eq:gammas}: the basis functions are single peaked with equally spaced maxima and variances that increase linearly with distance from the origin, although they have Gaussian rather than exponential tails.  These functions would make a good choice for two-sided data on an unbounded domain in which most of the data fall close to the origin, or they could be shifted $x \to x-x_0$ for use with data that are concentrated around some other value~$x_0$.  Figure~\ref{fig:gaussians} shows these functions for $T=7$.

\begin{figure}
\begin{center}
\includegraphics[width=\columnwidth]{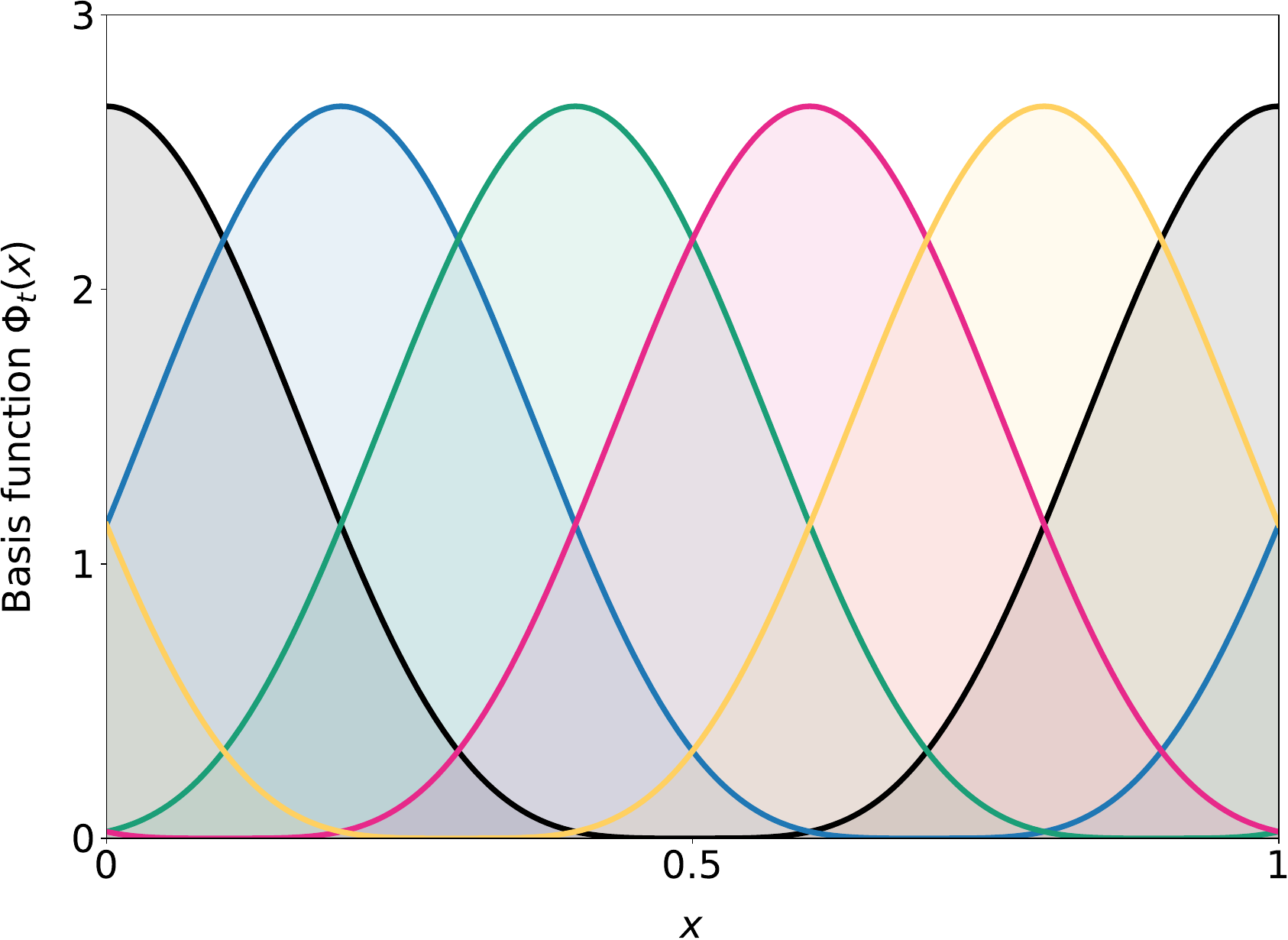}
\end{center}
\caption{The trigonometric polynomial basis functions of Eq.~\eqref{eq:trig} for $T=5$.  These functions are suitable for representing periodic probability distributions and are an analog of the Bernstein basis for functions on the line.}
\label{fig:trig}
\end{figure}

\subsection{Periodic distributions}
For data with a periodic distribution, such as angles or seasonal data, a periodic choice of basis functions is appropriate.  Various possibilities have been proposed, such as the von Mises distribution~\cite{EHP00}, which is a circular version of the normal distribution, or the trigonometric polynomial basis of Binette and Guillotte~\cite{BG22}, which is a circular version of the Bernstein basis.  The latter, for example, has the form
\begin{equation}
\Phi_t(x) = A \cos^{T-1} [\pi(x - t/T)],
\label{eq:trig}
\end{equation}
for $x\in[0,1]$, $t=0\ldots T-1$, and odd~$T$, with $A$ being a normalizing constant
\begin{equation}
A = T 2^{T-1} \Beta\bigl[ \tfrac12(T+1), \tfrac12(T+1) \bigr],
\end{equation}
where $\Beta(x,y)$ is the Euler beta function.  R\'oth~\etal~\cite{RJSH09} show that these functions form a complete basis for trigonometric polynomials of degree $(T-1)/2$.  A plot of the five $\Phi_t(x)$ for $T=5$ is shown in Fig.~\ref{fig:trig}.

\section{Proof of correctness for\\the Monte Carlo algorithm}
\label{app:mc}
In this appendix we give a proof that the Gibbs sampling algorithm of Section~\ref{sec:mc} does indeed sample correctly from the posterior distribution of Eq.~\eqref{eq:posterior}.  As with all Markov chain Monte Carlo methods, the proof requires us to demonstrate that the algorithm satisfies two criteria: ergodicity and detailed balance.

\subsection{Ergodicity}
Ergodicity is the requirement that every state of the system be accessible from every other by a finite sequence of Monte Carlo steps.  In the present case this requirement is trivially fulfilled, since our Monte Carlo steps can change any component or slot assignment and can add or delete components, allowing us to reach any state in a finite set of moves on a finite system.

\subsection{Detailed balance}
The more demanding part of the proof is the demonstration of detailed balance.  The condition of detailed balance says that the algorithm will sample correctly from the required distribution if the average rate, in equilibrium, of transitions between any two states~$\mu,\nu$ is the same in both directions.  If the equilibrium probability of a state~$\mu$ is $P(\mu)$ and the probability of a transition from $\mu$ to~$\nu$ is $P(\mu\to\nu)$, then the condition of detailed balance~is
\begin{equation}
P(\mu) P(\mu\to\nu) = P(\nu) P(\nu\to\mu),
\end{equation}
or, equivalently,
\begin{equation}
{P(\mu\to\nu)\over P(\nu\to\mu)} = {P(\nu)\over P(\mu)}.
\label{eq:detailed}
\end{equation}
Our proof of detailed balance follows the method outlined in~\cite{Newman25}.  We first rewrite the algorithm in a slightly different form for which the proof is more straightforward.  Then we show that this modified algorithm is equivalent to the one used in this paper.  The modified algorithm is as follows.

\begin{enumerate*}
\setlength{\itemsep}{4pt}
\item Choose a component~$r$ uniformly at random, then choose an observation~$i$ uniformly at random from component~$r$.
\item Remove $i$ from component~$r$.
\item If $i$ was the sole observation in component~$r$, so that its removal leaves the component empty, delete the component, relabel component~$k$ to be the new component~$r$, and decrease $k$ by~1.
\item Construct a set of $2k+1$ candidate states.
\begin{enumerate*}
\item Of these, $k$~states are ones in which $i$ is placed in an existing component~$s = 1\ldots k$.  Each of these states has an associated weight
\begin{equation}
w_\nu = {N-k\over k} P(k) \prod_j {\sum_{t=0}^{T_j-1} (m_{sjt}+1)\phi_{ijt}\over
         n_s+T_j},
\label{eq:proofw1}
\end{equation}
where $m_{sjt}$ and $n_s$ are the values after $i$ is removed from~$r$ but before it is placed in its new home.
\item The remaining $k+1$ candidate states are ones in which $i$ becomes the sole member of a new component $k+1$, then the labels of component $k+1$ and another component $s = 1\ldots k+1$ are exchanged.  (If $s=k+1$ then no exchange is performed.)  The weights for these states are
\begin{equation}
w_\nu = {k\over k+1} P(k+1) \prod_j {1\over T_j} \sum_{t=0}^{T_j-1} \phi_{ijt},
\label{eq:proofw2}
\end{equation}
where $k$ is the number of components before the new component is created.
\end{enumerate*}
\item Select a new state~$\nu$ for the system from the $2k+1$ candidates with categorical probabilities
\begin{equation}
p_\nu = {w_\nu\over\sum_\mu w_\mu}
\end{equation}
and update the system to that state.  If a new component is created, increase $k$ by~1.
\item Assign $i$ a new slot for each item~$j$ randomly from a categorical distribution with probabilities
\begin{equation}
p_j(t) = {(m_{sjt}+1) \phi_{ijt}\over \sum_{u=0}^{T_j-1} (m_{sju}+1) \phi_{iju}}
\label{eq:proofpjs1}
\end{equation}
in the general case, or
\begin{equation}
p_j(t) = {\phi_{ijt}\over\sum_{u=0}^{T_j-1} \phi_{iju}}
\label{eq:proofpjs2}
\end{equation}
for a newly created component.
\end{enumerate*}

To demonstrate that this algorithm satisfies detailed balance, there are various cases we need to address.  First, consider moves that do not change the number of components.  These come in several types.  The simplest are moves that remove an observation~$i$ from its component, then place it right back in the same component again.  In this case the only change in the state of the system comes from the reassignment of slots in step~6 above.  The probability $P(\mu\to\nu)$ of performing a move of this type is equal to the probability~$1/k$ of choosing the particular component~$r$, times the probability $1/(n_r+1)$ of choosing observation~$i$ from that component, times the probability $w_\nu/\sum_\mu w_\mu$ of choosing the particular move, and, finally, the probability of assigning new slots~$h_{ij}'$, where the prime denotes values in the new state.  (The factor of $1/(n_r+1)$ appears because, as above, $n_r$~denotes the value after $i$ is removed from~$r$.)  Altogether, we have
\begin{align}
P(\mu\to\nu) = {1\over k(n_r+1)}\, {w_\nu\over\sum_\mu w_\mu}
   \prod_j {(m_{rjh_{ij}'}+1) \phi_{ijh_{ij}'}\over
            \sum_t (m_{rjt}+1) \phi_{ijt}}.
\label{eq:forward}
\end{align}
The values of $k$ and $n_r$ are the same for the forward and backward moves, as are the values of all the~$w_\nu$.  Thus the ratio of transition probabilities in the two directions, in Eq.~\eqref{eq:detailed}, is
\begin{equation}
{P(\mu\to\nu)\over P(\nu\to\mu)}
  = \prod_j {(m_{rjh_{ij}'}+1) \phi_{ijh_{ij}'}\over
             (m_{rjh_{ij}}+1) \phi_{ijh_{ij}}}.
\label{eq:db1}
\end{equation}
Meanwhile the ratio of probabilities $P(\nu)/P(\mu)$ can be calculated from Eq.~\eqref{eq:posterior}.  Most of the factors in this expression are unchanged between states~$\mu$ and~$\nu$ and cancel out.  Once again letting $m_{rjt}$ and $n_r$ denote values after observation~$i$ is removed from component~$r$ but before it is reassigned to~$r$ again, the only factors that remain are
\begin{align}
{P(\nu)\over P(\mu)}
  &= {\prod_j m_{rjh_{ij}}! (m_{rjh_{ij}'}+1)!\over
      \prod_j (m_{rjh_{ij}}+1)! m_{rjh_{ij}'}!}
     {\prod_j \phi_{ijh_{ij}'}\over\prod_j \phi_{ijh_{ij}}} \nonumber\\
  &= \prod_j {(m_{rjh_{ij}'}+1) \phi_{ijh_{ij}'}\over
              (m_{rjh_{ij}}+1) \phi_{ijh_{ij}}},
\label{eq:db2}
\end{align}
which agrees with~\eqref{eq:db1} and detailed balance is established.

Another straightforward case is for Monte Carlo steps that remove the last observation~$i$ from a component~$r$, deleting that component, but then turn around and immediately place~$i$ back in a newly created component~$s$.  Again, most factors cancel from the probabilities and only those concerning the assignment of slots remain.  The details are essentially identical to those for the previous case, except that $m_{rjt}=m_{sjt}=0$ for all~$j,t$ by definition, since $i$ is the only observation that belongs to either component, and hence Eq.~\eqref{eq:db2} simplifies to
\begin{equation}
{P(\nu)\over P(\mu)}
  = \prod_j {\phi_{ijs_{ij}'}\over \phi_{ijs_{ij}}}.
\label{eq:db3}
\end{equation}
At the same time, the ratio $P(\mu\to\nu)/P(\nu\to\mu)$ also simplifies because of Eq.~\eqref{eq:proofpjs2}, and detailed balance is again established.

A more involved case occurs when an observation~$i$ is moved from component~$r$ to a different component~$s$ without the creation or deletion of either component but with the accompanying reassignment of slots for each item.  Again the probability of the forward move is given by Eq.~\eqref{eq:forward} and a similar expression applies for the reverse move, but now the factors of $n_r+1$~and $n_s+1$ do not cancel (because $r\ne s$), and although the value of $\sum_\mu w_\mu$ is the same in both directions (because the set of possible final states is the same), the individual weights~$w_\nu$ are different, and we get
\begin{align}
&{P(\mu\to\nu)\over P(\nu\to\mu)}
  = {(n_s+1)w_\nu\over(n_r+1)w_\mu} \nonumber\\
  &\qquad\times
   \prod_j {(m_{sjh_{ij}'}+1) \phi_{ijh_{ij}'}\over\sum_t (m_{sjt}+1) \phi_{ijt}}
   \prod_j {\sum_t (m_{rjt}+1) \phi_{ijt}\over(m_{rjh_{ij}}+1) \phi_{ijh_{ij}}}.
\end{align}
Using the value of $w_\nu$ from Eq.~\eqref{eq:proofw1}, many factors again cancel and we get
\begin{equation}
{P(\mu\to\nu)\over P(\nu\to\mu)} = {n_s+1\over n_r+1}
  \prod_j {(n_r+T_j) (m_{sjh_{ij}'}+1) \phi_{ijh_{ij}'} \over
  (n_s+T_j) (m_{rjh_{ij}}+1) \phi_{ijh_{ij}}}.
\label{eq:db4}
\end{equation}

\begin{widetext}
Meanwhile, from Eq.~\eqref{eq:posterior}, the ratio of probabilities in the two states is
\begin{align}
{P(\nu)\over P(\mu)} &= {n_r!(n_s+1)!
  \prod_j [m_{rjh_{ij}}! (m_{sjh_{ij}'}+1)!/(n_r+T_j-1)!(n_s+T_j)!]
          \phi_{ijh_{ij}'} \over (n_r+1)! n_s!
  \prod_j [(m_{rjh_{ij}}+1)! m_{sjh_{ij}'}!/(n_r+T_j)!(n_s+T_j-1)!]
          \phi_{ijh_{ij}}} \nonumber\\
  &= {n_s+1\over n_r+1}
  \prod_j {(n_r+T_j) (m_{sjh_{ij}'}+1) \phi_{ijh_{ij}'} \over
  (n_s+T_j) (m_{rjh_{ij}}+1) \phi_{ijh_{ij}}},
\end{align}
which agrees with~\eqref{eq:db4} and detailed balance is again established.
\end{widetext}

The final class of Monte Carlo steps we need to consider are ones where the number of components changes, because a component is created or deleted.  Without loss of generality, let us declare the forward move $\mu\to\nu$ to be the one in which a component is created by moving observation $i$ from component~$r$ to new component~$s$.  The probability of this move is the probability~$1/k$ of choosing component~$r$, times the probability~$1/(n_r+1)$ of choosing observation~$i$, times the probability $w_\nu/\sum_\mu w_\mu$ of choosing the particular move, times the probability~\eqref{eq:proofpjs2} of assigning new slots~$h_{ij}'$.  Putting these together, we have
\begin{equation}
P(\mu\to\nu) = {1\over k(n_r+1)}\,{w_\nu\over\sum_\mu w_\mu}
  \prod_j {\phi_{ijh_{ij}'}\over\sum_t \phi_{ijt}}.
\end{equation}
For the reverse move, in which $i$ is removed from $s$ and the component deleted, the probability of the move is equal to the probability~$1/(k+1)$ of choosing component~$s$ (because there are now $k+1$ components), times the probability $w_\mu/\sum_\mu w_\mu$ of choosing the move back to state~$\mu$, times the probability of assigning the original slots~$h_{ij}$.  This gives
\begin{equation}
P(\nu\to\mu) = {1\over k+1}\, {w_\mu\over\sum_\mu w_\mu}
   \prod_j {(m_{rjh_{ij}}+1) \phi_{ijh_{ij}}\over\sum_t (m_{rjt}+1) \phi_{ijt}}.
\end{equation}
Note that there is no factor of $1/(n_s+1)$ in this expression because $n_s=0$ by definition.

Now the ratio of rates in the forward and backward directions is
\begin{equation}
{P(\mu\to\nu)\over P(\nu\to\mu)}
  = {k+1\over k(n_r+1)} {w_\nu\over w_\mu}
     \prod_j {\phi_{ijh_{ij}'}\over\sum_t \phi_{ijt}}
     {\sum_t (m_{rjt}+1) \phi_{ijt}\over(m_{rjh_{ij}}+1) \phi_{ijh_{ij}}}.
\end{equation}

\begin{widetext}
Using Eqs.~\eqref{eq:proofw1} and~\eqref{eq:proofw2} for the weights~$w_\nu$ and~$w_\mu$, this now becomes
\begin{align}
{P(\mu\to\nu)\over P(\nu\to\mu)}
  &= {k+1\over k(n_r+1)}
    {[k/(k+1)] P(k+1) \prod_j (1/T_j) \sum_t \phi_{ijt} \over
     [(N-k)/k] P(k) \prod_j \sum_t (m_{rjt}+1) \phi_{ijt}/(n_r+T_j)}
    \prod_j {\phi_{ijh_{ij}'}\over\sum_t \phi_{ijt}}
     {\sum_t (m_{rjt}+1) \phi_{ijt}\over(m_{rjh_{ij}}+1) \phi_{ijh_{ij}}} \nonumber\\
  &= {k\over N-k} {1\over(n_r+1)} {P(k+1)\over P(k)}
     \prod_j {n_r+T_j\over T_j} {\phi_{ijh_{ij}'}\over(m_{rjh_{ij}}+1)\phi_{ijh_{ij}}}.
\label{eq:db5}
\end{align}
At the same time, from Eq.~\eqref{eq:posterior}, the ratio of probabilities for the two states is
\begin{align}
{P(\nu)\over P(\mu)} &= {P(k+1) k!(N-k-1)! n_r!
  \prod_j [(T_j-1)!]^2/(n_r+T_j-1)!\,T_j!\,m_{rjh_{ij}}!\,\phi_{ijh_{ij}'} \over
  P(k) (k-1)!(N-k)! (n_r+1)! \prod_j (T_j-1)!/(n_r+T_j)! (m_{rjh_{ij}}+1)! \phi_{ijh_{ij}}} 
  \nonumber\\
  &= {k\over N-k} {1\over(n_r+1)} {P(k+1)\over P(k)}
     \prod_j {n_r+T_j\over T_j} {\phi_{ijh_{ij}'}\over (m_{rjh_{ij}}+1) \phi_{ijh_{ij}}},
\end{align}
which is equal to~\eqref{eq:db5} and hence, once again, detailed balance is established.
\end{widetext}

This demonstrates that detailed balance is obeyed for all moves.  For the final step of our proof, we observe that the moves that create a new component $s=1\ldots k+1$ in step~4(b) of the algorithm are all equivalent up to a label permutation and hence, if we don't care about such permutations, we can lump them all together and represent them with a single move that creates a new component~$k+1$ and carries $k+1$ times the weight, Eq.~\eqref{eq:proofw2}, of each individual move:
\begin{equation}
w_\mu = k P(k+1) \prod_j {1\over T_j} \sum_{t=0}^{T_j-1} \phi_{ijt},
\end{equation}
as in Eq.~\eqref{eq:w2}.  This is the algorithm given in the main text and the one we use in our calculations.  Lumping states together like this means that when the algorithm performs a move followed by its reverse move, we may not end up back in the state we started with but in an equivalent state with permuted labels.  This has no effect on the division of the observations into components and slots, or on our estimate of the number of components, but it saves us some time and complexity in the algorithm.

This completes the proof of correctness for our algorithm.

\section{Data sets}
\label{app:data}
In this appendix we give additional details on the data sets used in our calculations.

\subsection{Italian wine}
This widely studied data set describes chemical and photometric measurements of 178 Italian wines made from three different grapes: barolo, grignolino, and barbera.  The original data set, due to Forina~\etal~\cite{FACU86}, contained 27 variables for each wine, but a smaller set containing 13 of the variables, compiled by Aeberhard~\etal~\cite{ADC94}, has been widely analyzed in the machine learning community, and this is the version we use also.  The data are available from many sources, for example the UC Irvine Machine Learning Repository at \texttt{https://archive.ics.uci.edu/dataset/109/wine}.  The 13 variables, as reported by Aeberhard~\etal, are:
\begin{enumerate*}
\item Alcohol content
\item Malic acid content
\item Ash content
\item Alcalinity of ash
\item Magnesium content
\item Total phenols
\item Flavanoids
\item Nonflavanoid phenols
\item Proanthocyanins
\item Color intensity
\item Hue
\item OD280/OD315 of diluted wines
\item Proline
\end{enumerate*}

\subsection{Cancer biomarkers}
This data set comes from Debernardi~\etal~\cite{Debernardi20} and is freely available at \texttt{https://doi.org/10.1371/ journal.pmed.1003489.s009} for download.  The data consists of anonymized measurements of four urinary biomarkers for pancreatic cancer in 590 patients, including patients both with and without cancer.  The four biomarkers are:
\begin{enumerate*}
\item Creatinine
\item Lymphatic vessel endothelial hyaluronan receptor~1 (LYVE1)
\item Regenerating family member 1 beta (REG1B)
\item Trefoil factor 1 (TFF1)
\end{enumerate*}
All four have non-negative values measured in ng/ml, but are rescaled in our analysis to have mean~$\frac12$ and arbitrary units.  The data set also includes basic demographics for the patients (age, sex), disease diagnosis (no pancreatic disease, benign disease, cancer), plus brief additional diagnostic data.  None of these latter variables are used as inputs to our analysis, but the diagnosis variable is used as ground truth for assessing the classification performance of the method.

\subsection{Personality test}
These data are from an online personality test fielded by the Open Source Psychometrics Project and are available online at \texttt{https://www.kaggle.com/datasets/ tunguz/big-five-personality-test}.  The test asked participants to judge how well each of 50 different statements described themselves, on a scale of 1 (strongly disagree) to 5 (strongly agree).  The statements are listed in Table~\ref{tab:big5} in the order in which they appeared on the test.

\begin{table*}
\setlength{\tabcolsep}{6pt}
\begin{center}
\begin{tabular}{llll}
\scshape{ext1} & I am the life of the party & \scshape{ext6} & I have little to say \\
\scshape{est1} & I get stressed out easily & \scshape{est6} & I get upset easily \\
\scshape{agr1} & I feel little concern for others & \scshape{agr6} & I have a soft heart \\
\scshape{csn1} & I am always prepared & \scshape{csn6} & I forget to put things back in their proper place \\
\scshape{opn1} & I have a rich vocabulary & \scshape{opn6} & I do not have a good imagination \\[1ex]

\scshape{ext2} & I don't talk a lot & \scshape{ext7} & I talk to a lot of different people at parties \\
\scshape{est2} & I am relaxed most of the time & \scshape{est7} & I change my mood a lot \\
\scshape{agr2} & I am interested in people & \scshape{agr7} & I am not really interested in others \\
\scshape{csn2} & I leave my belongings around & \scshape{csn7} & I like order \\
\scshape{opn2} & I have difficulty understanding abstract ideas & \scshape{opn7} & I am quick to understand things \\[1ex]

\scshape{ext3} & I feel comfortable around people & \scshape{ext8} & I don't like to draw attention to myself \\
\scshape{est3} & I worry about things & \scshape{est8} & I have frequent mood swings \\
\scshape{agr3} & I insult people & \scshape{agr8} & I take time out for others \\
\scshape{csn3} & I pay attention to details & \scshape{csn8} & I shirk my duties \\
\scshape{opn3} & I have a vivid imagination & \scshape{opn8} & I use difficult words \\[1ex]

\scshape{ext4} & I keep in the background & \scshape{ext9} & I don't mind being the center of attention \\
\scshape{est4} & I seldom feel blue & \scshape{est9} & I get irritated easily \\
\scshape{agr4} & I sympathize with others' feelings & \scshape{agr9} & I feel others' emotions \\
\scshape{csn4} & I make a mess of things & \scshape{csn9} & I follow a schedule \\
\scshape{opn4} & I am not interested in abstract ideas & \scshape{opn9} & I spend time reflecting on things \\[1ex]

\scshape{ext5} & I start conversations & \scshape{ext10} & I am quiet around strangers \\
\scshape{est5} & I am easily disturbed & \scshape{est10} & I often feel blue \\
\scshape{agr5} & I am not interested in other people's problems \phantom{xxxxx} & \scshape{agr10} & I make people feel at ease \\
\scshape{csn5} & I get chores done right away & \scshape{csn10} & I am exacting in my work \\
\scshape{opn5} & I have excellent ideas & \scshape{opn10} & I am full of ideas \\
\end{tabular}
\end{center}
\caption{The fifty statements used for the Big~5 personality test, listed in the order in which they appeared on the test.  Participants were asked to rate their agreement with each statement on a scale of 1 to~5, with 1 meaning strongly disagree and 5 meaning strongly agree.  The question IDs listed in the first and third columns encode the purpose of the questions, which are intended to probe participants standing with respect to the Big~5 personality traits.  The codes stand for extroversion ({\scshape ext}), emotional stability ({\scshape est}), agreeableness ({\scshape agr}), conscientiousness ({\scshape csn}), and openness ({\scshape opn}).}
\label{tab:big5}
\end{table*}

Each of the 50 questions was intended to probe the participant's standing with respect to one of the ``Big~5'' personality traits of extroversion, neuroticism (also called emotional stability), agreeableness, conscientiousness, and openness.  Questions were ask in groups of five, with one question in each group addressing each of the Big~5 traits, always with the traits in the order above.

The entire data set consists of responses from almost a million participants.  Many of these, however, are incomplete, with responses to one or more questions missing, and these we removed from the data.  We then took the first 10\,000 of the remaining participants for the data set we analyze.

\end{document}